\documentclass[useAMS,usenatbib]{mn2e}
\usepackage{graphicx}
\def\eq#1{\begin{equation} #1 \end{equation}}

\def\mic              {\hbox{$\mu{\rm m}$}}

\title[Panchromatic Properties of SDSS Galaxies]{Panchromatic Properties of 99,000 Galaxies 
 Detected by \\ SDSS, and (some by) ROSAT, GALEX, 2MASS, IRAS, \\ GB6, FIRST, NVSS and WENSS Surveys}
\author[M. Obri\'{c} et al.]{M. Obri\'{c}$^{1,2}$,  \v{Z}. Ivezi\'c$^{2}$, P.N. Best$^{3}$, R.H. Lupton$^{4}$, C. Tremonti$^{6}$, J. Brinchmann$^{5}$, 
\newauthor %
M.A. Ag\"ueros$^{2}$, G.R. Knapp$^{4}$, J.E. Gunn$^{4}$, C.M. Rockosi$^{7}$, D. Schlegel$^{4}$, D. Finkbeiner$^{4}$,
\newauthor %
M. Ga\'{c}e\v{s}a$^{13,10}$, V. Smol\v{c}i\'{c}$^{12,10}$, S.F. Anderson$^{2}$, W. Voges$^{8}$, M. Juri\'{c}$^{4}$,
\newauthor %
R.J. Siverd$^{4}$, W. Steinhardt$^{4}$, A.S. Jagoda$^{4}$, M.R. Blanton$^{9}$, D.P. Schneider$^{11}$\\
$^{1}$Kapteyn Astronomical Institute, University of Groningen, P.O. Box 800, Groningen 9700 AV, The Netherlands\\
$^{2}$University of Washington, Department of Astronomy, Box 351580, Seattle, WA 98195-1580, USA \\
$^{3}$Institute for Astronomy, Royal Observatory Edinburgh, Blackford Hill, Edinburgh EH9 3HJ, UK\\
$^{4}$Princeton University Observatory, Peton Hall, Princeton, NJ 08544-1001, USA \\
$^{5}$Max-Planck-Institute f\"ur Astrophysik, D-85748, Garching, Germany\\
$^{6}$Hubble Fellow, University of Arizona, Steward Observatory, 933 N. Cherry Ave., Tucson, AZ 85721, USA\\
$^{7}$University of California--Santa Cruz, 1156 High St., Santa Cruz, CA 95060, USA \\
$^{8}$Max-Planck-Institute f\"ur Extraterrestrische Physik, Karl-Schwarzschild-Str. 1, 
        Postfach 1317, D-85741 Garching, Germany\\
$^{9}$The New York University, Physics Department, 4 Washington Place, New York, NY 10003, USA \\
$^{10}$University of Zagreb, Physics Department, Bijeni\v{c}ka cesta 32, 10000 Zagreb, Croatia\\
$^{11}$Department of Astronomy and Astrophysics, 525 Davey Laboratory, University Park, PA 16802, USA\\
$^{12}$Max-Planck-Institut f\"ur Astronomie, K\"onigstuhl 17, Heidelberg, D-69117, Germany\\
$^{13}$University of Connecticut, Physics Department, 2152 Hillside Road, Storrs, CT 06269-3046, USA}

\begin{document}
\pagerange{\pageref{firstpage}--\pageref{lastpage}} \pubyear{200?}
\maketitle
\label{firstpage}
\begin{abstract}
We discuss the panchromatic properties of 99,088 galaxies selected from the
Sloan Digital Sky Survey Data Release 1 ``main'' spectroscopic 
sample (a flux-limited sample for 1360 deg$^2$). These galaxies are
positionally matched to sources detected by ROSAT, GALEX, 2MASS, IRAS,
GB6, FIRST, NVSS and WENSS. The matching fraction varies from
$<1$\% for ROSAT and GB6 to $\sim$40\% for GALEX and 2MASS. 
In addition to its size, the advantages of this sample are well
controlled selection effects, faint flux limits and the wealth of
measured parameters, including accurate X-ray 
to radio photometry, angular sizes, and optical spectra. We find strong 
correlations between the detection fraction at other wavelengths and 
optical properties such as flux, colors, and emission-line strengths. 
For example, $\sim$2/3 of SDSS ``main'' galaxies classified as AGN using
emission-line strengths are detected by 2MASS, while the corresponding fraction for 
star-forming galaxies is only $\sim$1/10. Similarly, over 90\% of galaxies
detected by IRAS display strong emission lines in their optical spectra, compared
to $\sim$50\% for the whole SDSS sample. Using GALEX, SDSS, and 2MASS data, we 
construct the UV-IR broad-band spectral energy distributions for various types 
of galaxies, and find that they form a nearly one-parameter family. For example, 
the SDSS $u$- and $r$- band data, supplemented with redshift, can be used to 
``predict'' $K$-band magnitudes measured by 2MASS with an rms scatter of only 
0.2 mag. When a dust content estimate determined from SDSS spectra with the aid
of models is also utilized, this scatter decreases to 0.1 mag and can be fully
accounted for by measurement uncertainties. We demonstrate that this interstellar 
dust content, inferred from optical SDSS spectra by Kauffmann et al. (2003a), is indeed 
higher for galaxies detected by IRAS and that it can be used to ``predict'' measured 
IRAS 60 $\mu$m flux density within a factor of two using
only SDSS data. We also show that the position of a galaxy in the
emission-line-based Baldwin-Phillips-Terlevich diagram is correlated
with the optical light concentration index and $u-r$ color determined 
from the SDSS broad-band imaging data, and discuss changes in the morphology of 
this diagram induced by requiring detections at other wavelengths. Notably, we find 
that SDSS ``main'' galaxies detected by GALEX include a non-negligible fraction 
(10-30\%) of AGNs, and hence do not represent a clean sample of starburst galaxies.
We study the IR-radio correlation and find evidence that its slope may be different
for AGN and star-forming galaxies and related to the $H_\alpha/H_\beta$ line
strength ratio. 
\end{abstract}


\section{Introduction}

The study of global galaxy properties has been recently invigorated by modern
sensitive large-area surveys across a wide wavelength range. The Sloan Digital Sky
Survey (SDSS, York et al. 2000, for more details see Appendix A1) stands
out because it has already provided near-UV to near-IR five-color
imaging data and high-quality spectra (R$\sim$1800) for over 500,000
galaxies. The ``main'' spectroscopic galaxy sample is defined by a
simple r-band flux limit (Strauss et al. 2002), and will include close to
1,000,000 galaxies. 

A number of detailed galaxy studies based on SDSS data have already been published.
Strateva et al. (2001) and Shimasaku et al. (2001) demonstrated a tight correlation
between the $u-r$ color, concentration of the galaxy's light profile, and morphology. 
Blanton et al. (2003) presented the SDSS galaxy luminosity function, and Kauffmann 
et al. (2003ab) determined and analyzed stellar masses and star-formation histories 
for 100,000 SDSS galaxies.

In addition to ``stand-alone'' studies based on only SDSS data, SDSS can be used as a
cornerstone for panchromatic studies of galaxies aided by recent surveys at
 wavelengths outside the optical range (0.3-1 \mic). The special role of
 SDSS in such studies is due to its rich optical information, in
 particular high-quality spectra and photometry. Nevertheless, galaxies emit a
 substantial fraction of their bolometric flux outside the wavelength
 range accessible to SDSS. For example, in starburst and Seyfert 2 galaxies
the mid/far-IR wavelength range is the most important contributor to the bolometric 
flux, e.g. Schmitt et al. (1997). Information obtained by other surveys offers important 
observational constraints for models of galaxy formation and evolution. 

Numerous studies that utilize SDSS and surveys at other wavelengths have already been 
published. For example, Finlator et al. (2000) analyzed the properties of point sources 
detected by SDSS and Two Micron All Sky Survey (2MASS), and Ivezi\'{c} et al. (2001a)
discussed the colors and counts of SDSS sources detected by SDSS, 2MASS, and FIRST surveys.
Ivezi\'{c} et al. (2002) cross-correlated SDSS and the survey Faint Images of
the Radio Sky at Twenty Centimeters (FIRST), and analyzed
the optical and radio properties of quasars and galaxies. Bell et al. (2003) used
SDSS and 2MASS data to estimate the baryonic mass functions of galaxies, and Anderson 
et al. (2003) studied the properties of AGN galaxies
detected by SDSS and ROSAT. Best et al. (2005ab) studied radio
galaxies, Chang et al. (2005) analyzed the SDSS-2MASS colors of
elliptical galaxies, and Goto (2005) and Pasquali, Kauffmann \& Heckman (2005)
studied the optical properties of SDSS galaxies detected by IRAS. A detailed
analysis of rest-frame colors in the Str\"{o}mgren system synthesized from 
SDSS spectra was presented by Smol\v{c}i\'{c} et al. (2006). They found
that the galaxy distribution in the resulting color-color diagrams forms
a very narrow locus with a width of only 0.03 mag. This finding agrees well
with the conclusion by Yip et al. (2004), based on a principal component analysis 
of SDSS spectra, that galaxy spectra can be described by a small number of 
eigenspectra.  

Here we cross-correlate the catalog of galaxies from SDSS Data Release 1
(Abazajian et al. 2003) with catalogs of sources detected by ROSAT (X
ray), GALEX (UV), 2MASS (near-IR), IRAS (mid/far-IR), GB6 (6 cm), FIRST
(20 cm), NVSS (20 cm), and WENSS (92 cm). References and a description
of each survey are listed in Appendix A. The panchromatic galaxy samples
discussed here are $\sim$10-100 times larger than those used in older
pre-SDSS studies. In addition, they are selected by simple flux limits,
and benefit from a wealth of accurately measured parameters including
X-ray to radio photometry, angular sizes, and optical spectra. The main aim of this
paper is to quantify the fraction and basic properties of SDSS ``main''
galaxies detected by other surveys using a uniform approach for all
analyzed surveys. However, due to the size and quality of the resulting
samples, even a simple, preliminary analysis presented here is
sufficient to yield a wealth of additional results. 

We describe our matching and analysis methods in Section 2. In Section 3 we
discuss the detection fraction of SDSS galaxies by other surveys, and in Section 
4 we present a preliminary analysis of some panchromatic properties of galaxies
in our sample. We discuss and summarize our results in Section 5.

\section{     Matching and Analysis Methods          } 

\begin{table*}
  \centering
  \label{matchtab}
   \caption{Catalogs, their wavelength range, matching radius, total number
     of matches (for SDSS total number of galaxies), false match probability
     for the adopted matching radius, and, in the bottom table, matching 
     fractions for all galaxies and for each galaxy class. 
     False match probabilities are computed from the source density in the 
     matched catalogs, and are consistent with the random matching rate when 
     SDSS positions are offset by 1 deg in declination.
     The matching fractions in the bottom table are corrected for the
     difference in area covered by each catalog and the area
     covered by SDSS DR1. Emission/No emission tags in the bottom table
     refer to whether or not the emission lines were detected in a
     galaxy (see text). The bottom row lists the surface density of each
     galaxy subsample in the SDSS DR1 catalog.}
   \begin{tabular}{@{}ccccccc@{}}
    \hline
    \hline
     Catalog     & Wavelength  & Matching distance [''] & All [absolute \#] & False match probability [\%] & False associations [\%] \\
    \hline
     SDSS        & NUV-NIR     &   -    & 99088 &  -      &  -      \\
     ROSAT RASS  & X-ray       &   30   &   267 &  0.056  &  8.9    \\
     GALEX       &  UV         &    6   &   866 &  0.796  &  1.9    \\
     2MASS XSC   & near-IR     &   1.5  & 19184 &  0.002  &  $<1$   \\
     IRAS  FSC   & far-IR      &   30   &  1736 &  0.013  &  $<1$   \\
     GB6         & 6 cm        &   20   &   132 &  0.003  &  14     \\
     FIRST       & 20 cm       &    3   &  3402 &  0.037  &  1.0    \\
     NVSS        & 20 cm       &   15   &  3478 &  0.291  &  8.3    \\
     WENSS       & 92 cm       &   20   &   363 &  0.227  &  9.1    \\
    \hline
    \hline
    Catalog & All [\%] & No emission [\%] & Emission [\%] & AGN [\%] & SF [\%] & Unknown [\%] \\
    \hline
     ROSAT RASS        &  0.63 &   0.63&  0.46 &  0.87 &  0.24 &  0.23   \\
     GALEX             & 42.0  &  24.1 & 67.7  & 27.9  & 82.1  & 93.3    \\
     2MASS XSC         & 38.1  &  35.7 & 39.6  & 63.8  & 10.7  & 34.0    \\
     IRAS  FSC         &  1.77 &   0.14&  2.95 &  3.01 &  1.73 &  3.54   \\
     GB6               &  0.22 &   0.15&  0.09 &  0.21 &  0.02 &  0.02   \\
     FIRST             &  3.86 &   2.76&  4.68 &  8.06 &  0.98 &  3.40   \\
     NVSS              &  3.51 &   2.85&  3.52 &  5.02 &  1.66 &  3.25   \\
     WENSS             &  2.50 &   3.23&  1.40 &  2.46 &  0.49 &  1.01   \\
    \hline
     SDSS [\#/deg$^2$] & 72.86 & 39.38 & 33.48 & 11.81 & 7.54  & 13.44   \\
    \hline
    \hline

 \end{tabular}
\end{table*}

There are 99,825 unique galaxies in the Sloan Digital Sky Survey Data Release 1 
``main'' spectroscopic sample\footnote{The recent SDSS Data Release 4 contains spectra
for 565,715 galaxies, see www.sdss.org}, a sample limited by Petrosian magnitude, 
$r_{Pet}<17.77$ and covering 1360 deg$^2$ (for more detailed description see Stoughton et al. 2002 and 
Strauss et al. 2002). We further restrict the sample by requiring the redshifts to lie in the 
range $0.01 \leq z \leq 0.30$ and obtain the sample of 99,088 galaxies analyzed here. For each 
galaxy, SDSS provides numerous properties measured from 5-color imaging data, 
such as astrometry, photometry, and morphological information, as well as high-quality
spectra. In addition to standard spectroscopic parameters automatically measured by 
the spectroscopic pipeline, we also utilize emission line measurements described 
by Kauffmann et al. (2003a). We emphasize that the SDSS astrometry is very accurate 
($\sim$0.1 arcsec, Pier et al. 2002), which significantly simplifies the matching algorithm. 

For each SDSS galaxy, we search for the two nearest neighbors in each of
the eight catalogs. We accept the nearest neighbor as a true association if 
its distance is smaller than the catalog-dependent matching radius listed in Table 1. 
The matching radius for each catalog was determined by analyzing the distribution
of distances between the quoted position in the catalog and the SDSS position, and
corresponds to a $\sim3\sigma$ cutoff\footnote{The distance distribution
  for the SDSS-NVSS sample is better fit by a sum of two
  Gaussians. However, this behavior has no significant consequence for
  the matching completeness and contamination.}. Due to either high 
astrometric accuracy of other catalogs (e.g., 2MASS), or their low source surface 
density (e.g., IRAS), the matching contamination rate (fraction of false associations) 
is typically very low ($<1\%$) at non-radio wavelengths and $\sim 10\%$
for the four radio surveys, as implied by both the source density in the matched catalogs, 
and the matching rate when SDSS positions are offset by 1 deg in declination.
The fraction of cases where two sources in other catalogs are found within the
matching radius is typically small ($<1\%$); in these cases we simply take the nearest 
neighbor to represent the true association. This fraction is sufficiently low that none 
of the conclusions presented in this paper change when both neighbors are excluded.

Before we proceed with the discussion of matching rate for each catalog, we describe 
our analysis methods in the next two sections.

\subsection{The global optical properties of galaxies in the SDSS DR1 main spectroscopic sample}

The first step in analyzing galaxies detected at other wavelengths is to
compare their distribution in the optical parameter space to that for 
the whole SDSS sample. The parameters measured by by the SDSS
photometric pipeline {\it photo} (Lupton et al. 2002) are numerous
($\sim$100), and we limit our preliminary analysis to the distribution of 
galaxies in optical color-magnitude-redshift space.

\subsubsection{        Color-magnitude-redshift distributions          }
\label{cmz}

\begin{figure}
\centering
\includegraphics[bb=20 17 592 779, width=\columnwidth]{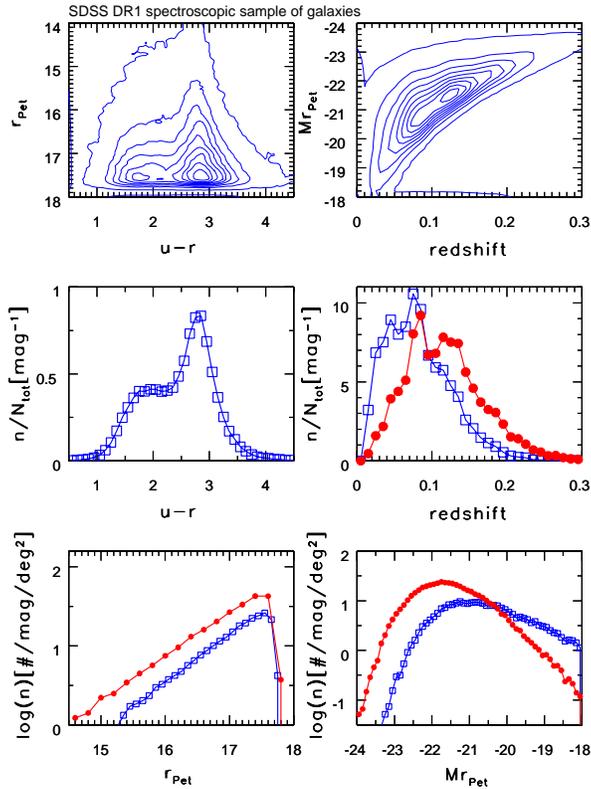}
\caption{Color-magnitude-redshift distributions for SDSS DR1 ``main'' galaxies 
($r_{Pet}<17.77$). 
The top two panels show the distribution of galaxies using linearly spaced contours, 
in steps of 10\%. The middle left panel shows the $u-r$ distribution of galaxies. 
Galaxies with $u-r < 2.22$ tend to be spiral/blue galaxies, and those with $u-r > 2.22$ 
elliptical/red galaxies (Strateva et al. 2001). Using this separation, the middle right 
panel shows the redshift (probability) distributions for blue (open squares) and red 
(dots) subsets. The same symbols are used to display their differential 
apparent magnitude (bottom left) and absolute magnitude (bottom right) distributions.
\label{SDSSdiags}}
\end{figure}

SDSS galaxies are not randomly distributed in the space spanned by apparent
(or absolute) magnitude, color, and redshift. As shown by Strateva et al. (2001),
and discernible in Fig.~\ref{SDSSdiags}, galaxies show a bimodal $u-r$ color
distribution (hereafter, optical SDSS colors are constructed using so-called ``model'' 
magnitudes; for details see Stoughton et al. 2002). Galaxies with $u-r < 2.22$ tend to 
be spiral galaxies, and those with $u-r > 2.22$ elliptical galaxies (see also Shimasaku 
et al. 2001 and Baldry et al. 2003). Because the spiral/blue galaxies tend to have lower 
luminosities than elliptical/red galaxies (bottom right panel), the
former are typically found at lower redshifts in the flux-limited SDSS
sample than are the latter. Blanton et al. (2003) give a detailed
analysis of the dependence of luminosity function on galaxy type. Note
that the ``features'' in the middle right panel of Figure 1 are due to
the large scale structure of galaxies. The differential number counts of
both color types are well described by $\log(n)=C+0.6\,r_{Pet}$ (see
also Yasuda et al. 2001). We will use diagrams such as
that shown in Fig.~\ref{SDSSdiags} to compare the distributions of
galaxies detected at other wavelengths to the distribution of all SDSS galaxies.

\subsubsection{    The distribution of emission-line galaxies in 
                        the Baldwin-Phillips-Terlevich diagram}
\label{Sbpt}

\begin{figure} 
\centering
\includegraphics[bb=72 84 540 708, width=\columnwidth]{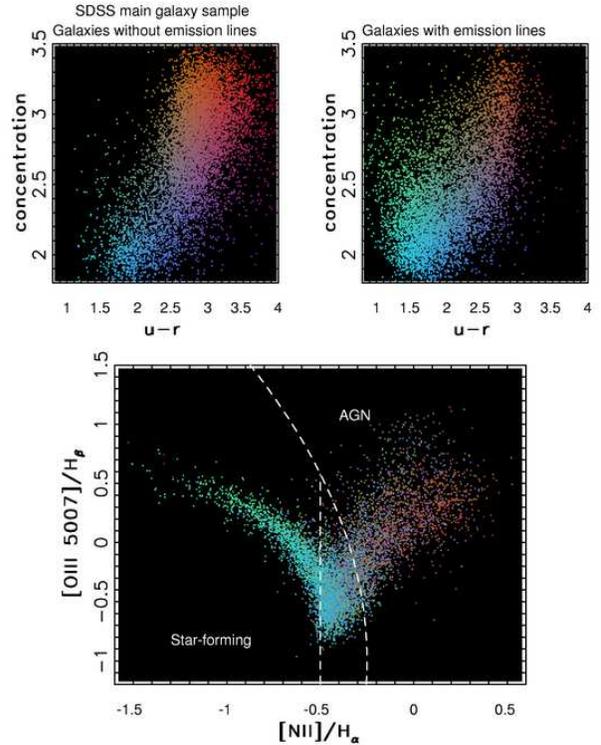}
\caption{The top two panels compare the distribution of galaxies without (left) and 
with (right) emission lines in the concentration index vs. $u-r$ diagram. 
The dots are two-dimensionally color coded according to their concentration index and 
$u-r$ color. The same color-coding scheme is used in the bottom panel, which shows the 
BPT diagram for emission-line galaxies (note that the line flux ratios are expressed
on a logarithmic scale). Emission-line galaxies can be separated into 
three groups according to their position in the BPT diagram: AGNs, star-forming, and 
``unknown'', using the separation boundaries outlined by the dashed lines. Note
the strong correlation between position in the BPT diagram and $u-r$.}
\label{BPTurC}
\end{figure}

\begin{figure} 
\centering
\includegraphics[bb=72 84 540 708, width=\columnwidth]{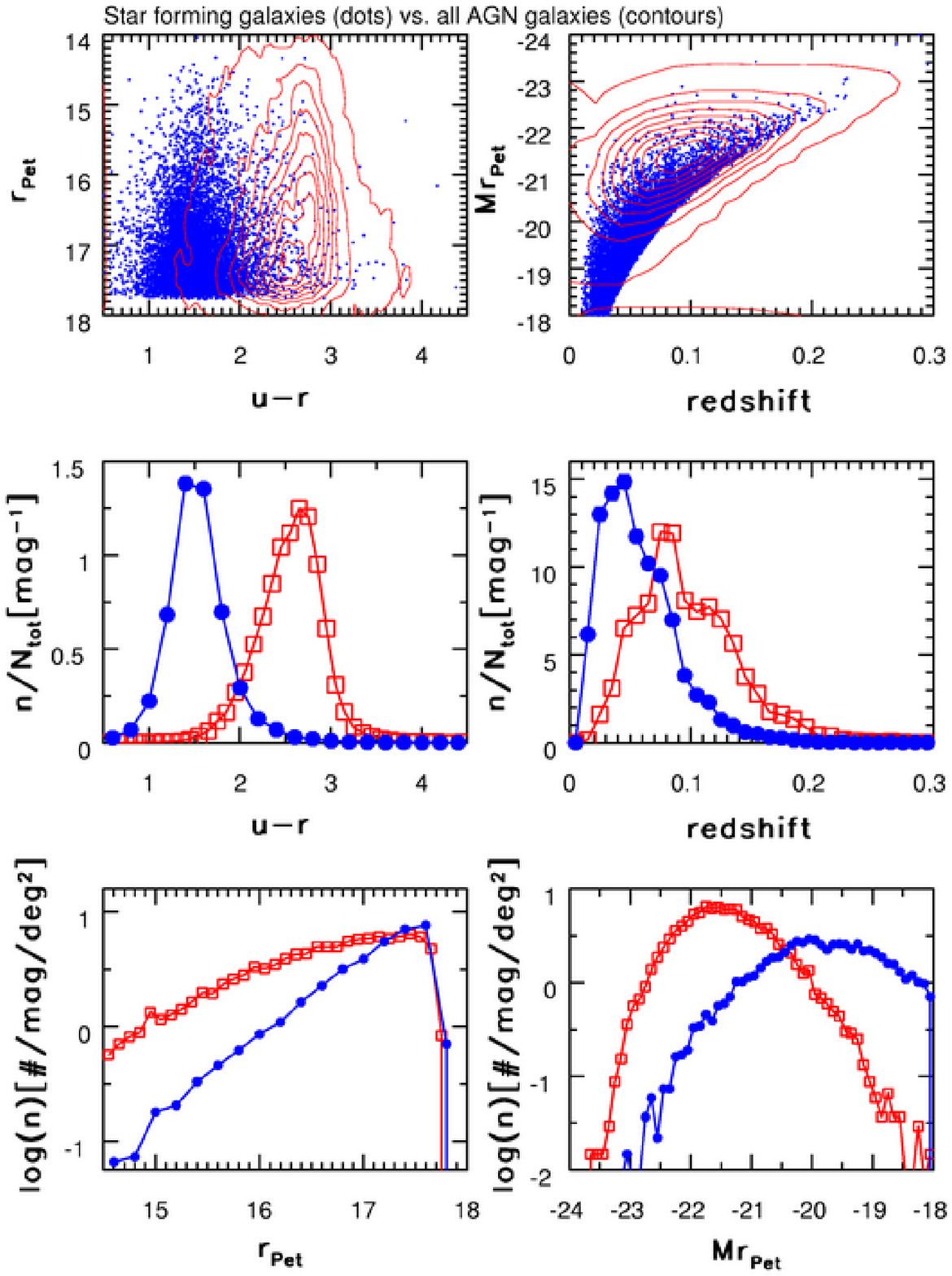}
\caption{The comparison of distributions of AGN (contours, squares) and
  star-forming (dots, filled circles) emission-line galaxies in
  magnitude-color-redshift space. Note remarkable differences in their
  broad-band imaging properties (especially $u-r$) showing the
  correlation of spectroscopic and photometric properties.}
\label{AGNvsSF}
\end{figure}

In addition to the overall comparison of galaxy distributions in diagrams shown in 
Fig.~\ref{SDSSdiags}, we analyze the behavior of three subsamples defined
by their emission line properties: galaxies without emission lines, and emission-line
galaxies separated into star-forming and AGNs. To classify a galaxy as 
an emission-line galaxy, we follow Kauffmann et al. (2003a) and require a $3\sigma$ 
significant detection of the $H_\alpha$, $H_\beta$, $[NII\,6583]$ and $[OIII\,5007]$  lines.
To classify emission-line galaxies as star-forming or AGN, we use the standard 
BPT diagram (Baldwin, Phillips \& Terlevich, 1981).

 The top two panels in Fig.~\ref{BPTurC}
compare the distribution of galaxies without and with emission lines in the 
concentration index vs. $u-r$ diagram. Galaxies without emission lines tend
to have larger concentration index and redder $u-r$ than galaxies with 
emission lines. The dots in these two panels are two-dimensionally color coded 
according to their concentration index and $u-r$ color. The same color-coding
scheme is used in the bottom panel, which shows the BPT diagram for emission-line
galaxies. {\it There is a strong correspondence between the position of a galaxy in the
BPT diagram and its position in the concentration index vs. $u-r$ diagram.}
Galaxies in the ``star-forming branch'' with small $[NII]/H_\alpha$ ratio, for 
a given $[OIII\,5007]/H_\beta$ ratio, have predominantly blue $u-r$ colors and small
concentration index, while AGNs have redder $u-r$ colors and large
concentration index. Furthermore, the distribution of {\it
  emission-line} galaxies in the BPT diagram is also correlated
with $u-r$ and concentration index.

In the subsequent analysis, we separate emission-line galaxies into three
groups according to their position in the BPT diagram: AGNs, star-forming, 
and ``unknown''. The adopted separation boundaries are shown by the dashed lines,
and are designed to produce robust clean samples of AGN and star-forming galaxies
(for alternative approaches see Hao et al. 2005, and references therein; for 
the aperture effects due to 3 arcsec fiber diameter see, e.g., Kauffmann et al. 
2003a and Kewley et al. 2005).
The ``unknown'' category is found at the join of the two branches, and it is not 
obvious from the displayed data to which class these galaxies belong. While their 
concentration index and $u-r$ color indicate that they may be star-forming galaxies, 
their IR colors and redshift distribution (see Section 4) suggest that they are 
more similar to AGN galaxies. Of course, it is possible that these objects are
star-forming galaxies that host an AGN.

The comparison of the distributions of  AGN and star-forming galaxies in 
magnitude-color-redshift space is shown in Fig.~\ref{AGNvsSF}. As discussed above,
the two types of galaxies, classified using emission lines, have remarkably
different $u-r$ distributions (see middle left panel). Furthermore, 
star-forming galaxies tend to have smaller luminosities than AGNs, and hence 
are observed at lower redshifts in the flux-limited SDSS sample. They also have 
very different differential number counts - the counts of AGN galaxies are 
flatter (d(logN)/dr $\sim$0.3) than those of star-forming galaxies (and those 
of the whole SDSS sample). For detailed studies of the optical
properties of star-forming and AGN galaxies, see e.g. Brinchmann et
al. (2004), Tremonti et al. (2004), Heckman et al. (2004), and
references therein.

\section{ What types of SDSS galaxies are detected at other wavelengths?}

The detection fraction of SDSS galaxies at other wavelengths is a strong function 
of optical properties such as flux, $u-r$, and emission-line strengths. 
In this analysis, we have taken into account the most important
selection effects, namely: observational biases caused by varying
survey depths, astrophysical effects such as intrinsically different color distributions
for different galaxy types, and K correction (Gunn \& Oke 1975) coupled with bias in redshift.
The size and quality of our sample, in addition to its well controlled selection criteria,
allow us to separate observational and astrophysical effects, and to study intrinsic
correlations amongst numerous measured galaxy properties.

The matching fraction varies from $<1$\% for ROSAT and GB6 to $\sim$40\% for 
GALEX and 2MASS (Table 1). We start the discussion with the 2MASS, continue
toward longer wavelengths, and then proceed from optical toward shorter wavelengths.

\subsection{                2MASS survey      }

\begin{figure} 
\centering
\includegraphics[bb=72 84 540 708, width=\columnwidth]{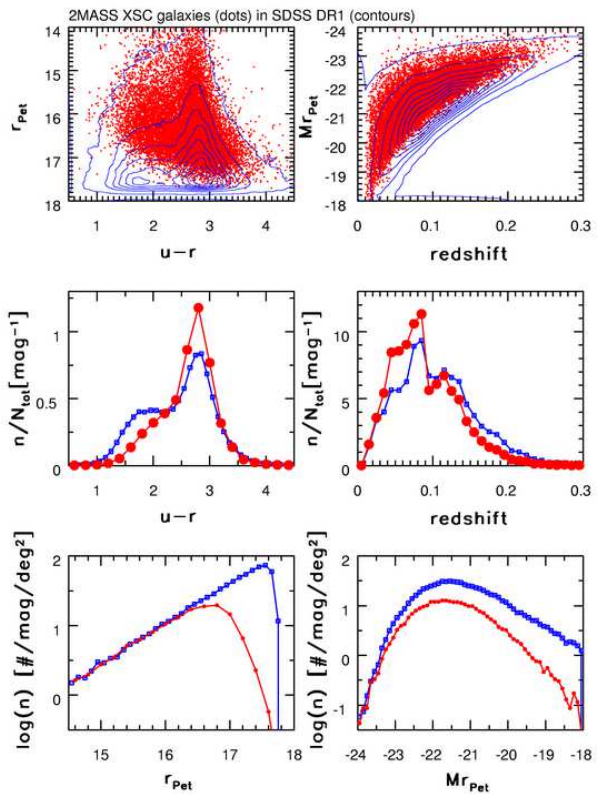}
\caption{The comparison of color-magnitude-redshift distributions for all SDSS 
DR1 ``main'' galaxies (contours, and histograms marked with small squares) and
those listed in 2MASS XSC (dots, and  histograms marked with large dots). 
2MASS XSC galaxies are biased towards red galaxies, larger luminosities,
and smaller redshifts. The 2MASS XSC catalog is essentially complete for galaxies
with $r_{Pet}<16.3$ (see bottom left panel).}
\label{2MASSopt}
\end{figure}

\begin{figure} 
\centering
\includegraphics[bb=72 84 540 708, width=\columnwidth]{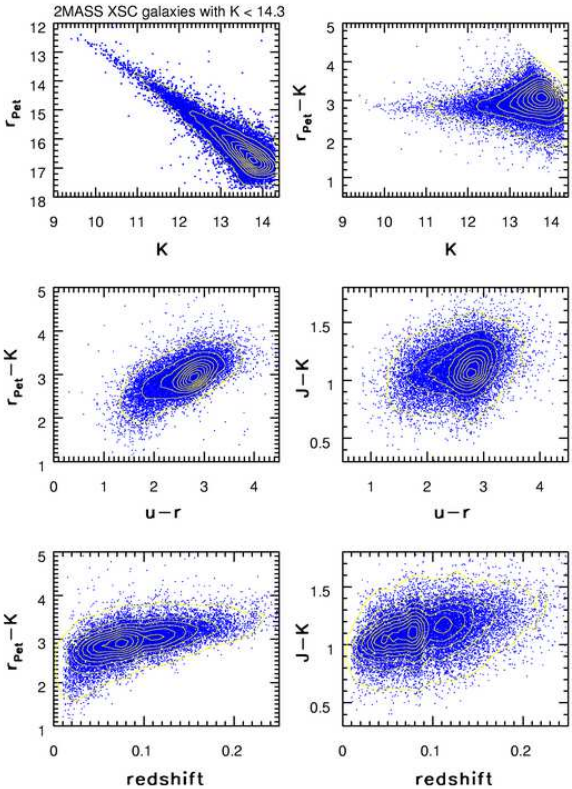}
\caption{The optical and IR colors of SDSS galaxies listed in 2MASS XSC.}
\label{2MASSanalyze}
\end{figure}

The distribution of SDSS galaxies detected by 2MASS\footnote{We analyze only resolved 
2MASS sources listed in the 2MASS XSC catalog; see Appendix A4 for more details.}
in the color-magnitude-redshift space is compared to the distribution of all 
SDSS galaxies in Fig.~\ref{2MASSopt}. The requirement that a galaxy is detected
and resolved by 2MASS (i.e., the XSC sample, see Appendix A4) introduces a
bias towards red galaxies and lower redshift as shown in the middle two
panels. The SDSS-2MASS XSC catalog is essentially 
complete\footnote{Here ``essentially complete'' implies a completeness of 
$\sim99\%$, as demonstrated by the direct comparison of the full SDSS and
2MASS overlap (see Finlator et al. 2000 and Ivezi\'{c} et al. 2001b).} for 
galaxies brighter than $r_{Pet}\sim16.3$ (bottom left panel, see also
McIntosh et al. 2005).

The color-dependent incompleteness of the 2MASS XSC catalog for galaxies with 
$r_{Pet}\ga16.3$ is due to the 2MASS faint limit, coupled with the
optical/IR color distribution of galaxies. We adopted $K=14.3$, which 
corresponds to a $\sim10\sigma$ detection; the $K$-band differential counts
of 2MASS XSC galaxies indicate that the catalog is complete to about
K=13.5. Fig.~\ref{2MASSanalyze} shows the optical/IR magnitude and color
distributions of SDSS-2MASS galaxies. We use 2MASS ``default''
magnitudes  (see Jarret et al. 2000) and do not correct for the
differences between AB (SDSS) and Vega (2MASS) magnitudes\footnote{For
  completeness, the AB-to-Vega offsets for SDSS bands are
  $m_0(u)=0.94$, $m_0(g)=-0.08$, $m_0(r)=0.17$, $m_0(i)=0.40$, and
  $m_0(z)=0.57$ (see http://www.sdss.org/ for details on the transformations)}. These
differences are $m_0=m_{AB}-m_{Vega}$, where $m_0(J)=0.89$,
$m_0(H)=1.37$, and $m_0(K)=1.84$ (see Finlator et al. 2000). Most galaxies have
$r_{Pet}-K$ in the range 2--3.5. Thus, the bluest galaxies 
are brighter than the 2MASS faint limit only if they have $r_{Pet}  < 16.3$. For 
$r_{Pet} > 16.3$ only galaxies with $r_{Pet}-K$ redder than $r_{Pet}-14.3$ 
are sufficiently bright in the $K$-band, and at $r_{Pet}=17.8$ practically 
no galaxies are listed in 2MASS XSC. This bias explains why the fraction of red 
galaxies in SDSS-2MASS sample is higher than among all SDSS galaxies (80\% vs. 66\%, 
with the blue/red separation defined by $u-r$=2.22). Since red galaxies tend to 
be more luminous than blue galaxies (e.g., Blanton et al. 2003), this color bias
also explains why 2MASS-SDSS galaxies are biased towards larger luminosities.

As the bottom two panels in Fig.~\ref{2MASSanalyze} demonstrate
$r_{Pet}-K$ and $J-K$ depend on redshift. This correlation (K correction),
coupled to the color effects discussed above, introduces a dependence of the
detection fraction on redshift. It is also an important effect to consider when
comparing the colors of various subsamples that may have different redshift distributions,
as we further discuss in Section~\ref{SagnVSsf}.

\subsubsection{   Predicting 2MASS K-band flux from UV/optical SDSS fluxes   }
\label{Kpredict}

The optical-IR color, $r_{Pet}-K$, is correlated with the UV-optical $u-r$ color,
as shown in the middle left panel in Fig.~\ref{2MASSanalyze}. This correlation 
indicates that it is possible to estimate the $K$-band flux using only
SDSS data, and is consistent with the fact that galaxies form a nearly one-dimensional 
sequence in various optical color-color diagrams constructed with SDSS data. 
The correlation among colors is especially tight for optical rest-frame colors, 
with a scatter of only $\sim$0.03 mag perpendicular to the locus (Smol\v{c}i\'{c} 
et al. 2006). The $r_{Pet}-K$ vs. $u-r$ correlation demonstrates that 
this one-dimensionality of broad-band galaxy spectral energy distributions (SEDs) extends 
to near-IR wavelengths.

In order to quantitatively assess to what extent near-IR flux is correlated
with optical fluxes, we determine the $K$-band flux from $K_{SDSS} =
r_{Pet} - (r-K)^\ast$, where $r_{Pet}$ is the SDSS $r$-band Petrosian magnitude
and $(r-K)^\ast$ is a best fit to the observed $r_{Pet}-K$ colors for SDSS/2MASS
galaxies sampled from SDSS. We use UV/visual fluxes ($u$- and $r$-bands) to 
fit the $r_{Pet}-K$ color because this is the ``hardest'' wavelength combination with most 
astrophysical implications. According to ``common wisdom'', such a relationship
should not be very accurate due to the effects of starbursts and dust
extinction. Predicting, for example, 2MASS $J$-band
flux from SDSS $z$-band flux is trivial because these two bands are
adjacent in wavelength space. We also take into account the K correction
(see the bottom left panel in Fig.~\ref{2MASSanalyze}), and fit the following functional form
\begin{equation}
 \label{Kpredeq}
      (r-K)^\ast = \\
             A+B\,(u-r)+C\,(u-r)^2+D\,(u-r)^3+E\,z_r+F\,z_r^2
\end{equation}
where $z_r$ is redshift. 

\begin{figure} 
\centering
\includegraphics[bb=90 280 532 729, width=\columnwidth]{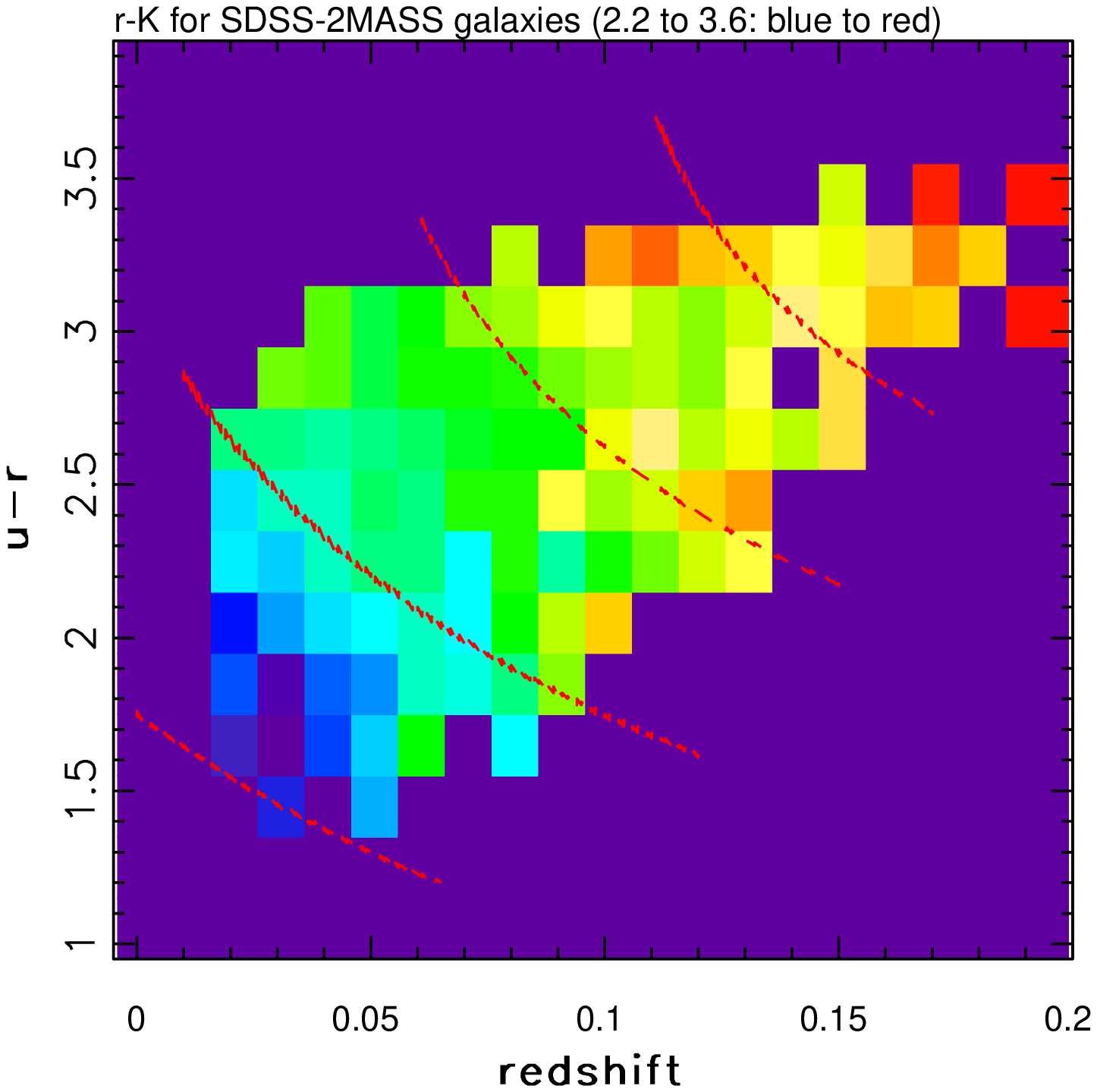}
\includegraphics[bb=90 280 532 729, width=\columnwidth]{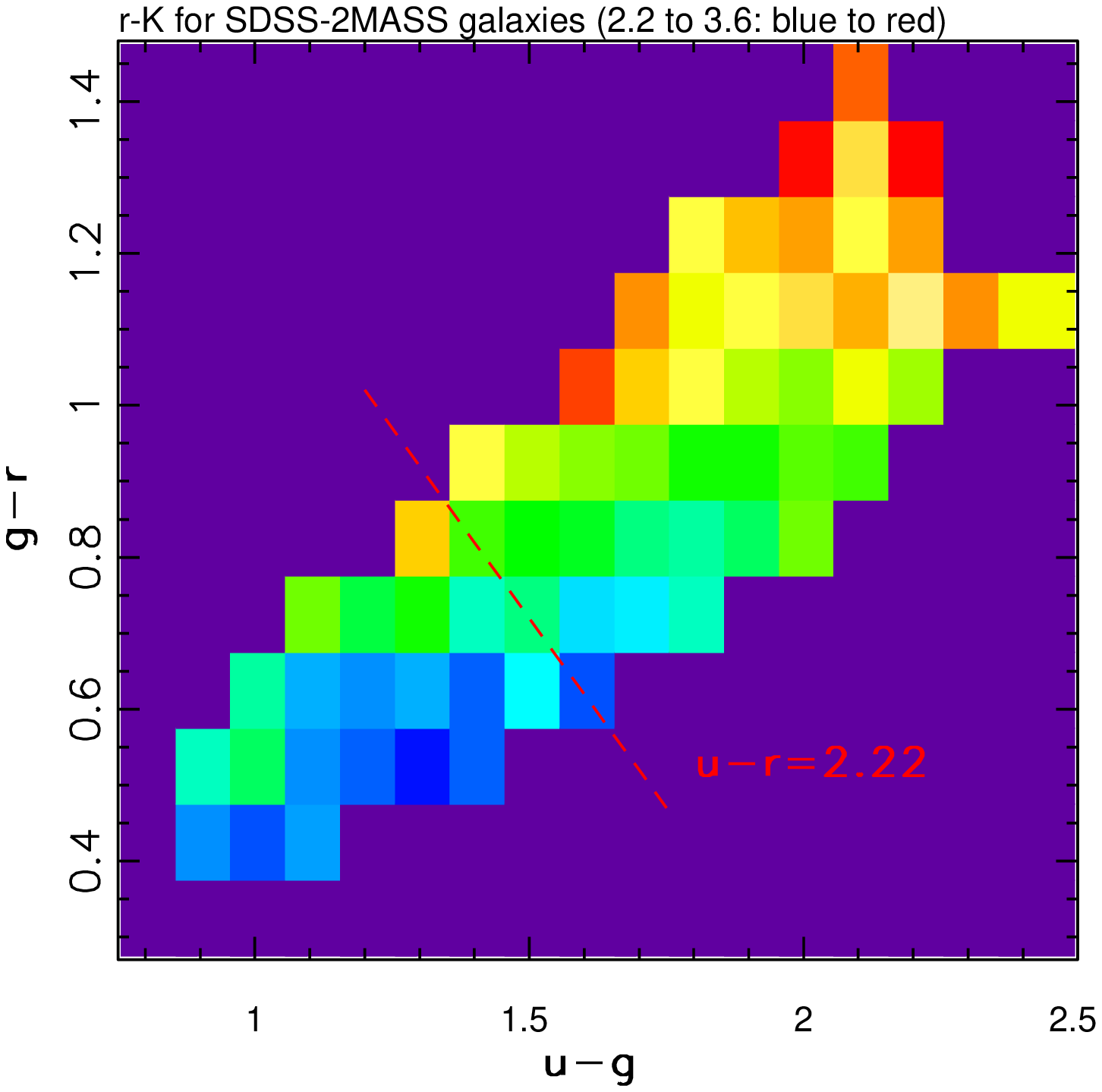}
\caption{The top panel shows the median $r-K$ color in small $u-r$ vs. redshift
bins. The median color is color-coded blue to red in the 2.2--3.6 range
with a linear stretch. Bins with fewer than 3 galaxies are shown with
magenta color. The distribution of the root-mean-scatter of $r-K$ color
per bin has a median of 0.2 mag and a width of 0.08 mag. The dashed
lines represent contours of constant $r-K$ (2.3, 2.7, 3.0 and 3.2) given
by eq.1. The bottom panel shows the median $r-K$ color in small $g-r$
vs. $u-g$ color bins, with analogous color coding. The distribution of the
root-mean-scatter of $r-K$ color per bin has a median of 0.23 mag and
a width of 0.10 mag. The dashed line shows the boundary between
blue and red galaxies ($u-r$=2.22) proposed by Strateva et al. (2001).}
\label{Kpred2D}
\end{figure}

The motivation for this functional form is the behavior of the $r-K$ color
shown in the top panel in Fig.~\ref{Kpred2D}. The $r-K$ color also seems to be a
well-defined function of the position in the $g-r$ vs. $u-g$ color-color
diagram, shown in the bottom panel in Fig.~\ref{Kpred2D}, but the scatter around
the median values in each bin is larger than for the $u-r$ vs. redshift diagram
(this is essentially due to larger uncertainty of photometric redshifts that
are implied by the position of a galaxy in the $g-r$ vs. $u-g$ diagram,
in comparison to spectroscopic redshifts). 

\begin{figure} 
\centering
\includegraphics[bb=72 84 540 708, width=\columnwidth]{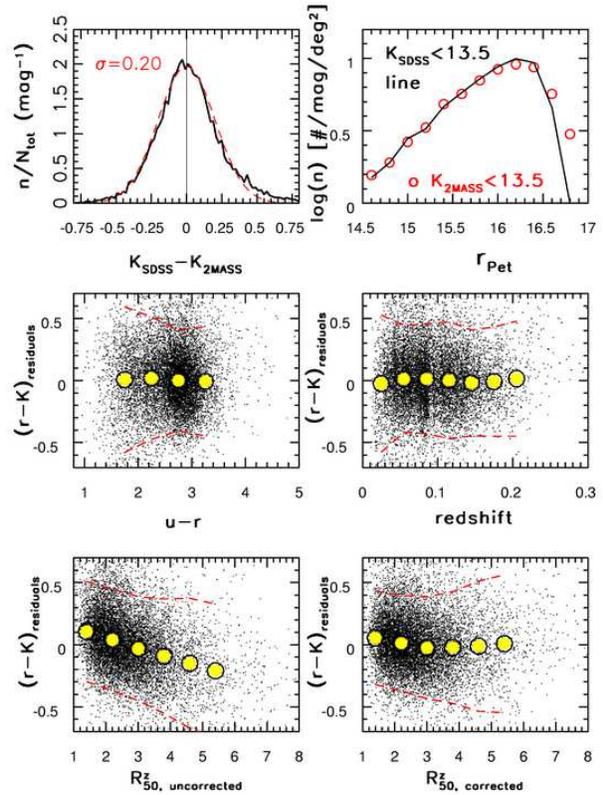}
\caption{The solid line in the top left panel shows the distribution of differences 
between SDSS-predicted and 2MASS-measured $K$-band flux, using eq. 1. The dashed line 
is a Gaussian with $\sigma=0.20$ mag. The right panel shows the $r$-band counts of 
SDSS-predicted (line) and 2MASS-measured (symbols) galaxies with $K<13.5$. The two
middle panels show the dependence of residuals between predicted and measured K-band
magnitudes on the $u-r$ color (left) and redshift (right). The bottom left panel
show the dependence of the same residuals on $R_{50}^z$, the radius enclosing
50\% of the Petrosian flux  in the $z$-band, before (left) and after (right) correcting
for this effect. In the lower four panels, large yellow circles represent
medians and 2$\sigma$ envelope is given by red dashed curves.}
\label{Kpred}
\end{figure}

With the best-fit values (A, B, C, D, E, F)= (1.115, 0.940, -0.165,
0.00851, 4.92, -9.10), this relation predicts 2MASS $K$-band magnitudes
with an rms scatter of only 0.20 mag. The residuals between the fitted and measured values
depend on neither color nor redshift, and are nearly Gaussian (see the
two middle panels in Fig.~\ref{Kpred}). However, there is a correlation between the residuals
and the optical galaxy size, parametrized by $R_{50}^z$, the radius enclosing
50\% of the Petrosian flux  in the $z$-band (for details see Stoughton et al. 2002, 
and Strauss et al. 2002).  To correct for these aperture and resolution effects, 
which presumably depend on galaxy profile\footnote{This assumption was recently
verified by Chang et al. (2005).}, or nearly equivalently on galaxy color
(Strateva et al. 2001), we add to the right hand side of the
eq.~\ref{Kpredeq} $\Delta(r-K)^\ast=(0.496-0.154\,R_{50}^z)$ for 
galaxies with $u-r<2.22$ and $\Delta(r-K)^\ast=(0.107-0.045\,R_{50}^z)$ for redder galaxies. 
This correction has a negligible effect on the rms scatter in the predicted K magnitude,
and only removes a correlation of $K_{SDSS}-K_{2MASS}$ residuals with galaxy size
(see the two bottom panels in Fig.~\ref{Kpred}).

The distribution of differences between the predicted and measured $K$-band magnitudes
is shown in Fig.~\ref{Kpred} (top left panel). The median residuals, as a function of $u-r$ and $z_r$, 
do not exceed 0.03 mag, and the rms scatter decreases to 0.15 mag at the bright end 
($K<12$). The top right panel in Fig.~\ref{Kpred} compares the differential
number counts as a function of $r_{Pet}$ for galaxies with $K<13.5$, where the 
latter condition is imposed using measured and predicted values. The good agreement
shows that predicted $K$-band flux is {\it not overestimated} for galaxies 
that are not in 2MASS XSC, and indicates that the proposed relations may be applicable
for galaxies fainter than the 2MASS faint cutoff. 

Given typical measurement errors in $u$, $r$, $R_{50}^z$  and $K$, we conservatively conclude 
that the true astrophysical scatter of $K$-band magnitudes predicted
from the blue part of the SED is not larger than $\sim$0.1
mag. Similarly, the relation $(J-K)=2.172\,z_r + 0.966$, where $z_r$ 
is redshift (see the bottom right panel in Fig.~\ref{2MASSanalyze}),
predicts the $J-K$ measured by 2MASS with an rms scatter of 0.11 mag
(0.07 mag at the bright end), and no significant residuals with respect
to $K$, $u-r$, and redshift; that is, the rest-frame $J-K$ color
distribution of {\it all} low-redshift galaxies is very narrow:
$\sim$0.1 mag. These tight correlations demonstrate the remarkable
one-dimensionality of galaxy spectral energy distributions from UV to IR
wavelengths. We further discuss the spectral energy distribution of
galaxies in Section~\ref{colorzsed}, and an improvement to the K band flux prediction
given by the eq.~\ref{Kpredeq} in Section~\ref{KpredImprove}.

\subsection{                IRAS FSC     }

\begin{figure} 
\centering
\includegraphics[bb=72 84 540 708, width=\columnwidth]{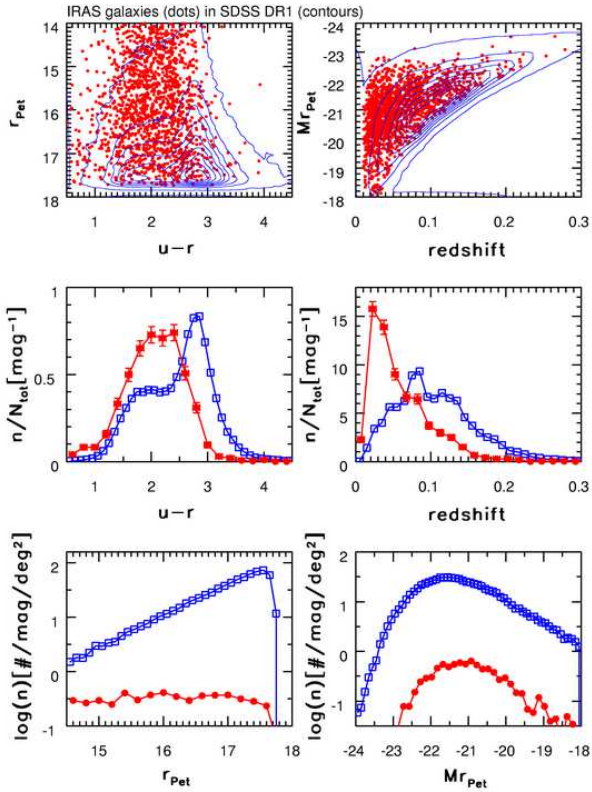}
\caption{Analogous to Fig.~\ref{2MASSopt}, except that here galaxies listed
in the IRAS FSC (dots, filled circles) are compared to the whole SDSS sample
(contours, squares). Galaxies detected by IRAS are strongly biased
towards blue galaxies and lower redshifts. The fraction of galaxies
detected by IRAS decreases from 22\% for $r_{Pet}=14.5$ to 1\% for 
$r_{Pet}=17.5$.}
\label{IRASopt}
\end{figure}

The distribution of SDSS galaxies detected by IRAS\footnote{We analyze the sources 
listed in the IRAS FSC catalog; see Appendix A5 for more details.} in 
color-magnitude-redshift space is compared to the distribution of all
SDSS galaxies in Fig.~\ref{IRASopt}. The requirement that a galaxy is
detected by IRAS introduces a strong bias towards optically blue
galaxies and lower redshift (the middle two panels). The majority of
these galaxies have emission lines and include both star-forming
and AGN galaxies, as we discuss in more detail in
Section~\ref{SagnVSsf}. The completeness of the IRAS FSC catalog depends strongly 
on $r_{Pet}$, and varies from 22\% for $r_{Pet}\sim14.5$ to 1\% for $r_{Pet}\sim17.5$.

\subsubsection{  The correlation between $u-m_{60}$ and galaxy dust content }
\label{SAz}

\begin{figure} 
\centering
\includegraphics[width=\columnwidth]{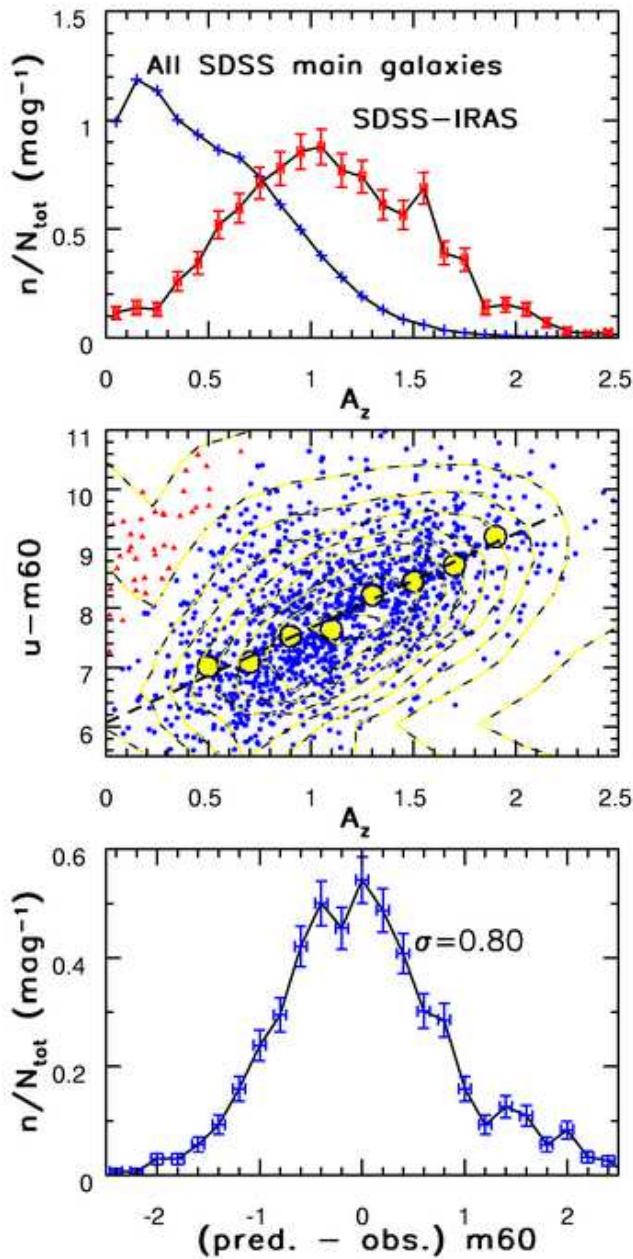}
\caption{The top panel compares the distributions of the $z$-band dust extinction, $A_z$, 
inferred from SDSS spectra by Kauffmann et al. (2003a), for all SDSS galaxies (pluses)
and for the subset detected by IRAS (squares with error bars). The middle panel shows the UV-far-IR 
color, $u-m_{60}$, as a function of $A_z$ (both symbols and contours). The large symbols 
mark median values of $u-m_{60}$ in $A_z$ bins in the 0.4--2.0 range. The dashed line 
is the best linear fit to these medians. The triangles, concentrated in the upper left
corner, mark the $\sim$3\% of galaxies that apparently do not follow this relation. The bottom panel
shows the distribution of differences between the 60 $\mu$m flux measured by IRAS,
and predicted values using the best-fit $u-m_{60}$ vs. $A_z$ relation,
and the $u$-band fluxes measured by SDSS. The rms of the distribution is marked in the panel
($\sigma$, in magnitudes).}
\label{IRASpred}
\end{figure}

Interstellar dust absorbs UV and optical radiation and re-emits it at
mid- and far-IR wavelengths. Hence, some degree of correlation is expected between
the far-IR-optical/UV colors and the amount of dust in a galaxy. 

Kauffmann et al.~(2003a) used the distribution of galaxies in the plane spanned
by the strength of the $H_\delta$ line and the 4000 \AA\ break ($D_{4000}$) to obtain
model-dependent estimates of stellar masses and dust content for SDSS galaxies. 
Given the position of a galaxy in the $H_\delta$--$D_{4000}$ plane, the most 
probable mass-to-light ratio is drawn from a model library. With the measured
luminosity, this ratio then yields stellar mass. The observed luminosity is
corrected for the dust extinction determined by comparing observed imaging
$g-r$ and $r-i$ colors to model-predicted colors (the latter do not include
the effects of dust reddening). The reddening correction needed to make
models agree with data is interpreted as an effective optical depth in the 
SDSS $z$-band, $A_z$, due to a galaxy's interstellar dust. Here we find, using the 
measured properties of galaxies detected by IRAS, {\it independent} support for the 
notion that these model-dependent estimates of $A_z$ are indeed related to the galaxy dust 
content.

The top panel in Fig.~\ref{IRASpred} compares the distributions of
$A_z$, determined by Kauffmann et al. (2003a), for all SDSS galaxies
(pluses) and for the subset detected by IRAS (squares with error
bars). Galaxies detected by IRAS have systematically higher values of $A_z$
than the full SDSS ``main'' galaxy sample. If values of $A_z$, determined using
{\it only} SDSS data, were not related to the dust content, there would be no 
systematic difference induced by requiring a detection by the fully {\it independent} 
IRAS survey.

Furthermore, we find a correlation between $u-m_{60}$ color ($m_\lambda$ are
IRAS measurements expressed as AB magnitudes, with $\lambda$=12, 25, 60, and 100
$\mu$m) and $A_z$. The small symbols in the middle panel in Fig.~\ref{IRASpred} 
show $u-m_{60}$ color as a function of $A_z$ for 1200 highly probable 
SDSS-IRAS identifications, selected from the full SDSS-IRAS sample by limiting
the maximum SDSS-IRAS distance to 20 arcsec. We first determine median values of 
$u-m_{60}$ in $A_z$ bins of range 0.4--2.0, and then fit a linear relation
to obtain
\begin{equation} 
      u-m_{60} = (6.0\pm0.2) + (1.64\pm0.2)\,A_z.
\end{equation}
We do not find significant differences in the best-fit relations fitted separately
to AGN and star-forming subsamples (classified using emission-line strengths). 
The adopted $A_z$ range excludes $\sim$3\% of the sample that has very small $A_z$ 
and $u-m_{60}$ about 2.5 mag redder than predicted by the above
relation (triangles in the upper left corner in the middle panel in
Fig. 8). It is not clear whether this 60$\mu$m excess is physical, or
due to random matches. In any case, the fraction of the excluded sources is
sufficiently small to have no effect on the overall correlation.

Using this relation and the measured SDSS $u$-band fluxes, we estimate the 60 $\mu$m
flux, and compare it to the measured values in the bottom panel in Fig.~\ref{IRASpred}. 
The IRAS 60 $\mu$m flux can be predicted within a factor of $\sim$2 (rms, or 0.8 mag) 
using {\it only} SDSS data. If, instead, the $A_z$ estimates are ignored, and the 
60 $\mu$m flux is estimated by assuming $u-m_{60}$ = 7.6 (the median value) for all 
galaxies, the rms scatter between the predicted and measured values becomes
1.41 mag. Hence, the $A_z$ estimates do contain information about the dust
content. 

We have also attempted to use the $H_\alpha/H_\beta$ line strength ratio as 
a proxy for effective dust extinction (e.g., see Moustakas, Kennicutt \& Tremonti
2005, and references therein). The values of $H_\alpha/H_\beta$  and $A_z$
determined by Kauffmann et al. (2003a) are well correlated. We find that
$H_\alpha/H_\beta$ for emission-line galaxies can be determined from $A_z$ with 
an rms scatter of 0.07 using the relationship 
\eq{
          {H_\alpha \over H_\beta} = 0.49 + 0.143*A_z.
}
This relationship maintains its accuracy when only radio-selected, IR-selected,
and subsamples separated into AGN and star-forming galaxies are considered.
Thus, the $u-m_{60}$ vs. $A_z$ and $H_\alpha/H_\beta$ vs. $A_z$ correlations
imply the existence of a $u-m_{60}$ vs. $H_\alpha/H_\beta$ correlation.

As expected, we find an overall correlation between $u-m_{60}$ and 
$H_\alpha/H_\beta$ ratio for SDSS-IRAS galaxies. However, it is not as strong 
as the $u-m_{60}$ vs. $A_z$ relation discussed above. In terms of the rms scatter 
between predicted and measured 60 $\mu$m flux, it predicts 60 $\mu$m magnitude 
within 1.2 mag, that is, not as well as when using $A_z$. This implies that $A_z$
values determined using SDSS spectra and sophisticated stellar population models
may be a better estimator of effective dust content than the straightforward 
application of the $H_\alpha/H_\beta$ line-strength ratio.

Although it is hard to estimate the errors in IRAS flux measurements without 
an independent data set, the radio-IR correlation discussed in \S~\ref{radioIR}
suggests that they are not larger than $\sim$0.4 mag, and therefore smaller
than the rms scatter of 0.82 mag between predicted and measured 60 $\mu$m fluxes
(the quoted formal IRAS FSC photometric errors are $\sim$0.2 mag). Thus,
it may be possible to further improve the prediction for far-IR flux by using additional 
SDSS measurements such as sizes and UV/optical colors. For example, the differences 
between predicted and measured 60 $\mu$m fluxes are somewhat correlated with 
$u-r$ color: the median value is $-0.1$ mag for galaxies with $u-r$ $<$ 2.22 and
0.2 mag for redder galaxies. A similar effect is seen when the sample is separated
into AGN and star-forming galaxies. We postpone such an analysis until
the larger samples needed for robust quantitative multi-dimensional analysis are constructed.

\subsection{   Radio surveys (GB6, FIRST, NVSS, WENSS)      }

The advent of modern sensitive large-area radio surveys (see Appendix A6 for 
brief descriptions and references), combined with an optical survey such as
SDSS, offers significantly larger, more diverse, and accurate samples of radio
sources with optical identifications than available until recently. Detailed
studies of SDSS sources detected by the FIRST and NVSS 20 cm surveys was presented by
Ivezi\'{c} et al. (2002) and Best et al. (2005ab). Here we extend their analysis to multiwavelength
radio observations by including data from the GB6 (6 cm) and WENSS (92 cm). 

\begin{figure} 
\centering
\includegraphics[bb=20 50 560 760, width=\columnwidth]{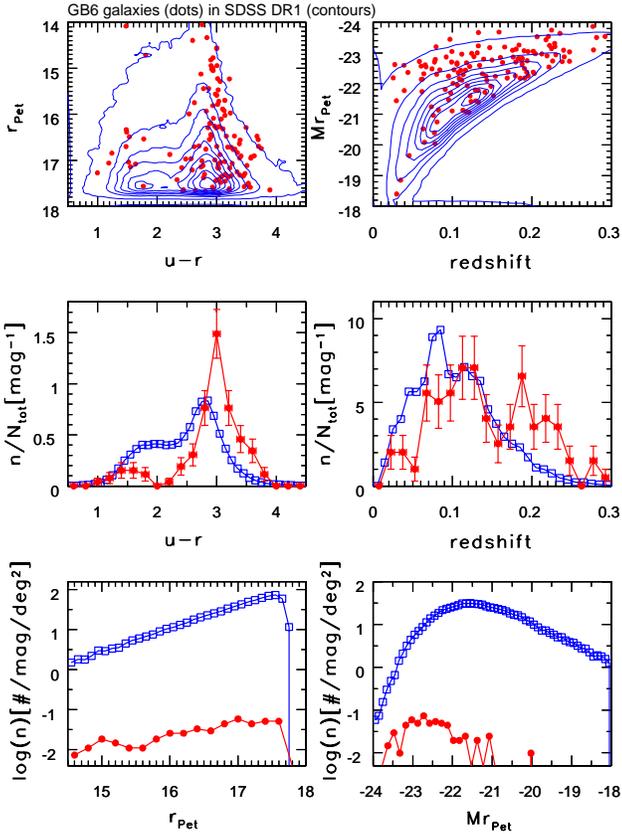}
\caption{Analogous to Fig.~\ref{2MASSopt}, except that here galaxies detected 
by GB6 (dots, filled circles) are compared to the whole SDSS sample
(contours, squares). Galaxies detected by GB6 are 
biased towards luminous red galaxies. The fraction of galaxies detected by GB6
is 0.22\%.}
\label{GB6}
\end{figure}

\begin{figure} 
\centering
\includegraphics[bb=20 50 560 760, width=\columnwidth]{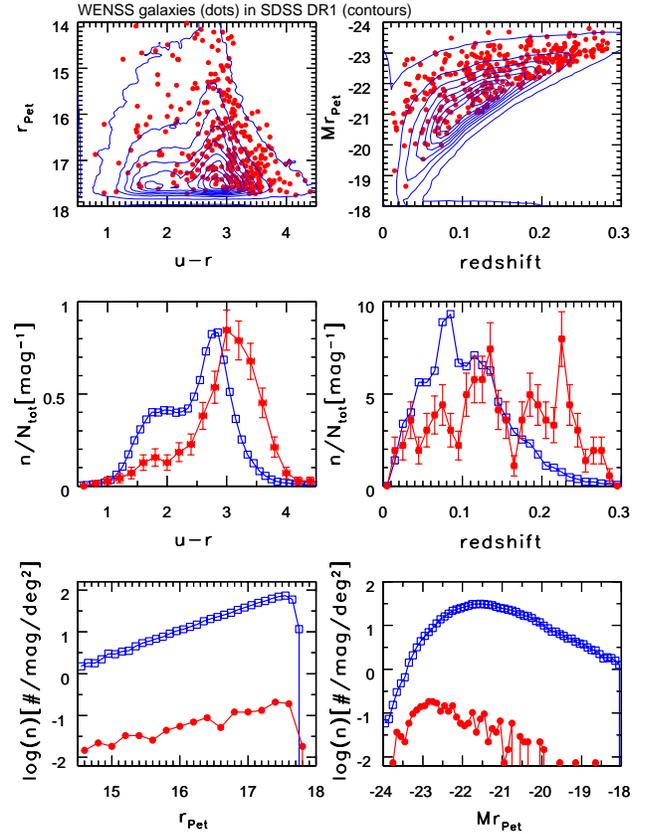}
\caption{Analogous to Fig.~\ref{2MASSopt}, except that here galaxies detected 
by WENSS (dots, filled circles) are compared to the whole SDSS sample
(contours, squares). Galaxies detected by WENSS are biased towards
luminous red galaxies. The fraction of galaxies detected by WENSS is 2.5\%.}
\label{WENSS}
\end{figure}

A summary of optical properties of SDSS galaxies detected by GB6 and WENSS
 is shown in Figs.~\ref{GB6} and \ref{WENSS}. Analogous diagrams for
galaxies detected by FIRST and NVSS can be found in Ivezi\'{c} et al.
(2002). The matching rate is the smallest for the GB6 catalog (0.22\%, see Table 1),
and the highest for FIRST (3.86\%). The difference in matching fractions for the two
20 cm surveys (FIRST and NVSS) is due to their different faint flux limits and
angular resolution. All four radio catalogs show similar distributions in 
color-magnitude-redshift space, despite the relatively large wavelength
coverage and varying angular resolution. Radio galaxies are
biased towards red, luminous galaxies and higher redshifts. Even when red
galaxies ($u-r>2.22$) are considered separately, their median $u-r$ color is
redder by about 0.3 mag for the radio-detected subsample than for the
whole red sample. However, this is simply a consequence of a bias in
redshift induced by requiring radio-detection coupled with the K
correction, as discussed by Ivezi\'{c} et al. (2002). When compared in a
small redshift range, {\it the radio-detected red galaxies 
have the same $u-r$ color distribution as red galaxies without radio detections}.

\begin{figure} 
\centering
\includegraphics[bb=20 290 560 740, width=\columnwidth]{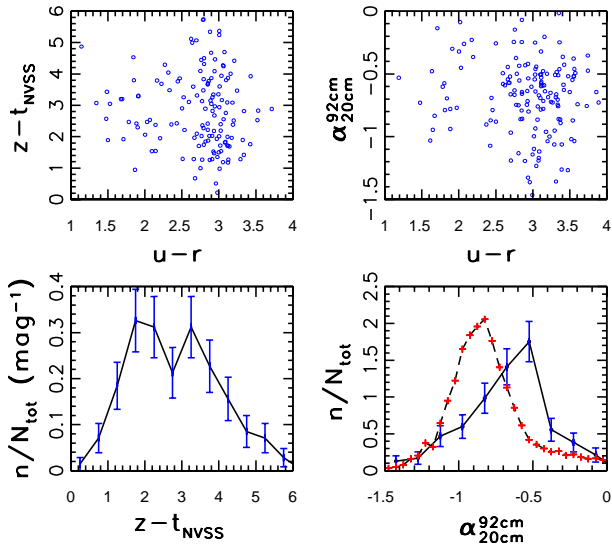}
\caption{The top left panel shows the radio-optical color (20 cm and the $z$-band) vs. 
$u-r$ color for a subsample of these galaxies with redshift in the range $0.10 \leq z \leq 0.14$, 
and the bottom left panel shows the distribution of their radio-optical color.
The top right panel shows the radio spectral slope ($\alpha$ for
$F_\nu \propto \nu^\alpha$ between 20 cm and 92 cm) vs. $u-r$ color for SDSS 
galaxies detected by NVSS and WENSS surveys. The bottom right panel
compares the distribution of the radio spectral slope for this sample
(histogram with error bars) to the distribution of the radio spectral slope 
for all radio sources detected by the NVSS and WENSS surveys (histogram with pluses).}
\label{radioPlot}
\end{figure}

This sample is sufficiently large to test whether the radio spectral 
slope is correlated with optical properties, such as $u-r$. 
Using NVSS and WENSS measurements, we compute the radio spectral slope 
between 20 cm and 92 cm, and find no correlation with the $u-r$ color 
(see the top right panel in Fig.~\ref{radioPlot}).  We find that the 
distribution of this spectral slope for ``main'' SDSS galaxies is different 
from the distribution for the full multiwavelength radio sample (bottom
right panel in Fig.~\ref{radioPlot}); the latter have a larger fraction of 
sources with ``steep'' spectra ($\alpha\sim -1$). This difference is probably 
caused by distant radio-galaxies that are not present in the ``main'' SDSS sample,
and by quasars. For further discussion of the distribution of galaxies and quasars 
in radio ``color-color'' diagrams, we refer the reader to Ivezi\'{c} et al. (2004a).

We also analyzed the radio-to-optical flux ratio as a function of $u-r$.
The top left panel in Fig.~\ref{radioPlot} shows the radio-optical color\footnote{Following
Ivezi\'{c} et al. 2002, we express all radio fluxes on AB magnitude
scale. We would like to apologize to radio astronomers, as
this seemed less of a problem than expressing SDSS and the UV-to-IR measurements in Janskys.} $z-t_{NVSS}$
as a function of $u-r$ color for a subsample of SDSS-NVSS-WENSS galaxies with redshift 
in the range $0.10 \leq z \leq 0.14$. We use the $z$-band because the 
dependence of the bolometric correction for galaxies on color is the smallest in 
this band (see Section~\ref{colorzsed}), and restrict the redshift range to minimize 
the effects of K correction. There is no discernible correlation between the
radio-optical and $u-r$ color. However, the measured distribution of the radio-optical 
color, shown in the bottom left panel, is subject to numerous selection effects
(such as multiple faint flux limits), and it is hard to uncover the intrinsic
distribution without detailed simulations, which will be attempted elsewhere.

\subsection{                 GALEX               }

\begin{figure} 
\centering
\includegraphics[bb=20 50 560 760, width=\columnwidth]{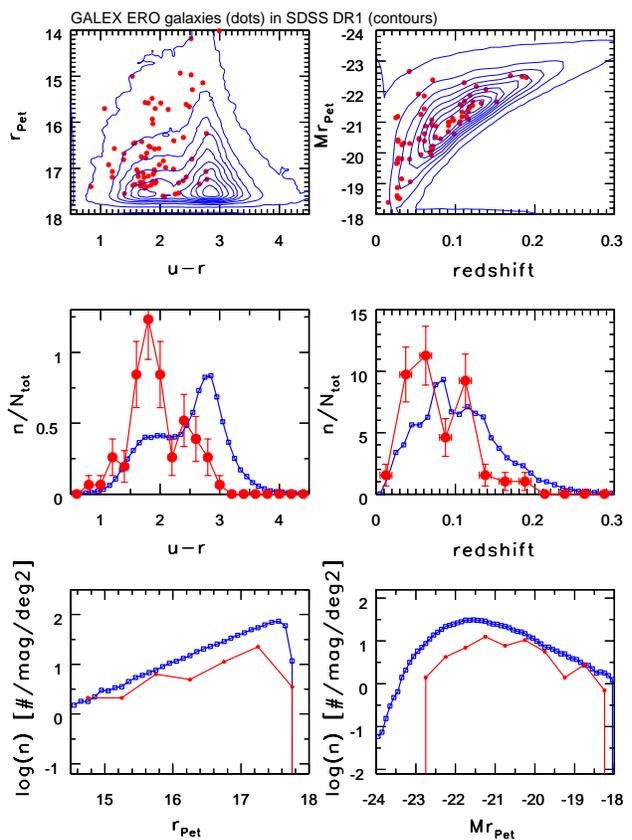}
\caption{Analogous to Fig.~\ref{2MASSopt}, except that here galaxies
  detected by GALEX (dots, filled circles) are compared to the whole
  SDSS sample (contours, squares). Galaxies detected by GALEX are biased
  towards blue galaxies, and are dominated (70\%) by emission line galaxies.
The latter include both star-forming and AGN ($\sim$10\%) galaxies
(compare the middle left panel to Fig.~\ref{AGNvsSF}). The fraction of SDSS ``main''galaxies 
detected by GALEX is $\sim$42\%, and approaches 100\% at the bright end.}
\label{GALEXopt}
\end{figure}

The distribution of SDSS galaxies detected by GALEX\footnote{We analyze sources 
listed in the GALEX Early Release Observations; see Appendix A3 for more details.} in 
color-magnitude-redshift space is compared to the distribution of all SDSS galaxies 
in Fig.~\ref{GALEXopt}. The requirement that a galaxy is detected by GALEX introduces 
a bias towards blue galaxies and lower redshift (the middle two panels). The fraction 
of SDSS ``main'' galaxies detected by GALEX\footnote{For an analysis of SDSS sources
detected by GALEX that is not limited to SDSS ``main'' galaxies, we refer the reader
to Ag\"ueros et al. (2005) and references therein.} is $\sim$42\%, and approaches 100\% 
at the bright end (the bottom left panel). The comparison of their $u-r$ distribution 
with those shown in Fig.~\ref{AGNvsSF} suggests that they are dominated by star-forming  
($u-r \la 2.2$) galaxies, but also include AGN ($2 < u-r < 3$) galaxies. The UV colors 
measured by GALEX support this conclusion (Ag\"ueros et al. 2005). Furthermore, the majority 
(70\%) of these galaxies have emission lines, and their distribution in the BPT diagram
(discussed in more detail in Section~\ref{multiBPT}) confirms that AGN emission, rather 
than starbursts, is the origin of UV flux in at least 10\% of SDSS-GALEX galaxies.

\subsection{                 ROSAT FSC         }

\begin{figure} 
\centering
\includegraphics[bb=20 50 560 760, width=\columnwidth]{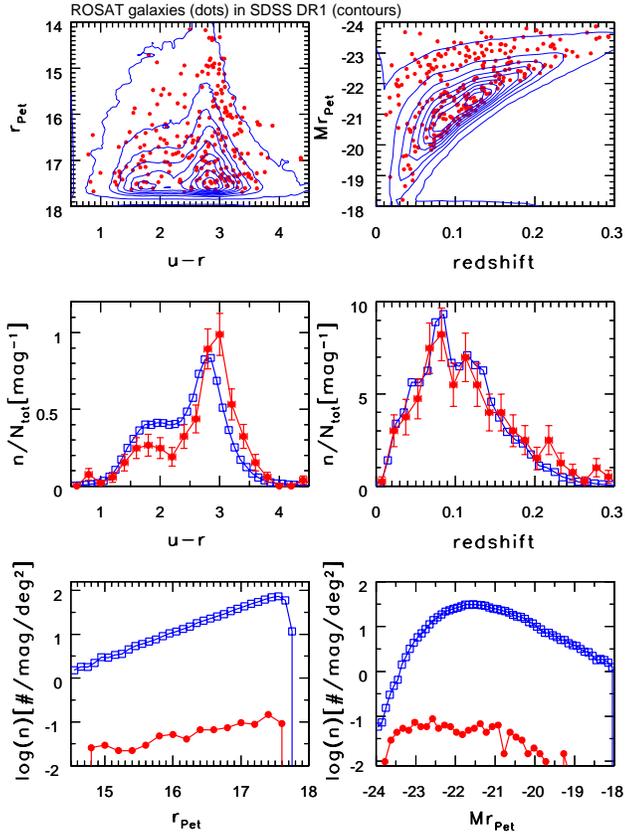}
\caption{Analogous to Fig.~\ref{2MASSopt}, except that here galaxies
  detected by ROSAT (dots, filled circles) are compared to the whole
  SDSS sample (contours, squares). Galaxies detected by ROSAT are biased
  towards red luminous galaxies. The fraction of galaxies detected by ROSAT
decreases from 0.6\% for $r_{Pet}=14.5$ to 0.2\% for $r_{Pet}=17.5$.}
\label{ROSATopt}
\end{figure}

Matching ROSAT X-ray detections (including both hard and soft X-ray data) to SDSS DR1 
optical counterparts produced a low (0.63\%)  matching
fraction. Color-magnitude-redshift diagrams (Fig.~\ref{ROSATopt}) reveal
a bias toward red galaxies, similar to that seen for radio surveys, but
without a redshift bias. The small sample size prevents more detailed analysis. 
We refer the reader to Anderson et al. (2003) for an analysis of SDSS sources 
detected by ROSAT that is not limited to SDSS ``main'' galaxies.

\section{       Panchromatic Properties of SDSS Galaxies            }

In this Section we combine the data from multiple surveys to construct and
compare the UV-IR spectral energy distributions (SEDs) for various subsamples of
galaxies, and analyze the changes in the BPT diagram induced by requiring 
detection at different wavelengths spanning the X-ray to radio range.

\subsection{   Dependence of colors on redshift  and  mean SEDs }
\label{colorzsed}

\begin{figure} 
\centering
\includegraphics[bb=72 224 550 708, width=\columnwidth]{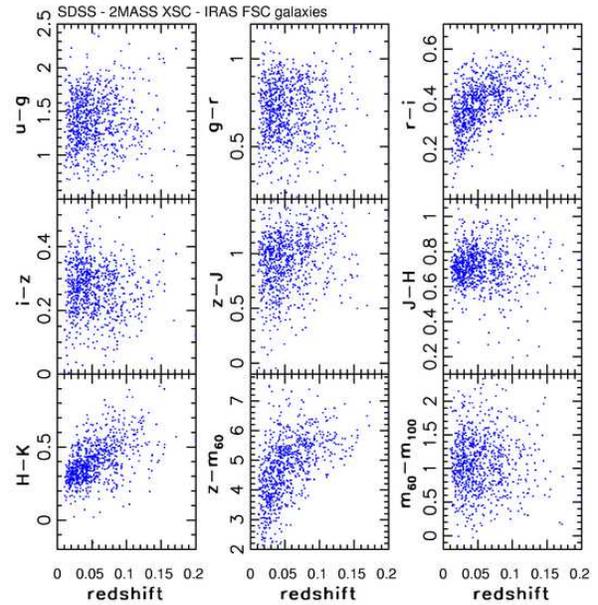}
\caption{The dependence of optical and infrared colors on redshift for SDSS 
galaxies listed in 2MASS XSC and IRAS FSC catalogs. Colors are constructed
with SDSS ``model'' magnitudes (AB system) and 2MASS XSC ``default'' 
magnitudes (Vega system). IRAS measurements at 60 $\mu$m and 100 $\mu$m
are expressed on AB system ($m_{60}$ and $m_{100}$).}
\label{Kcorr3IR}
\end{figure}

\begin{figure} 
\centering
\includegraphics[bb=70 87 542 709, width=\columnwidth]{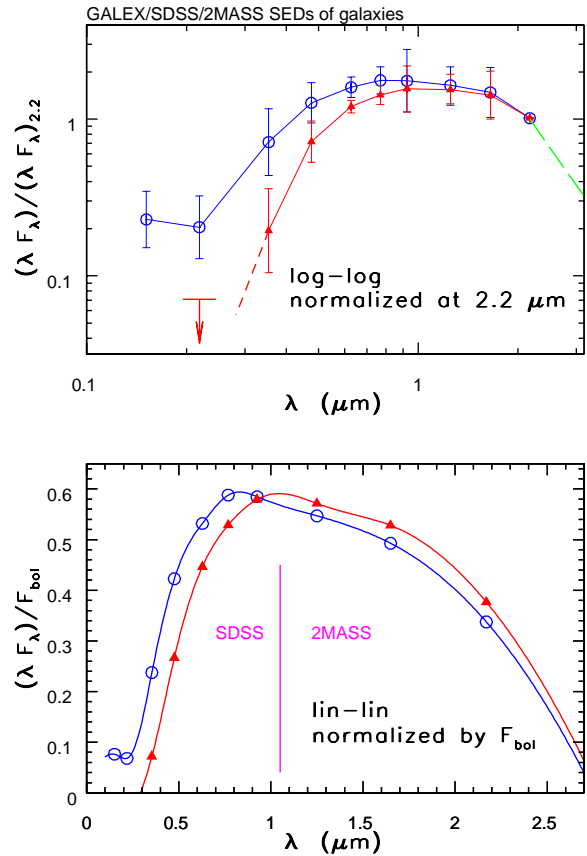}
\caption{The top panel compares the UV-to-near-IR SEDs for blue 
($u-r<2.22$, circles) and red ($u-r>2.22$, triangles) galaxies detected by
the GALEX, SDSS, and 2MASS surveys, and with redshifts in the range $0.03 \leq z \leq 0.05$.
The error bars indicate the root-mean-square color scatter (determined from the 
interquartile range). The data points are connected to guide 
the eye. The dashed line extending redwards from the $K$-band point (2.2 $\mu$m) 
is a Rayleigh-Jean extrapolation. The bottom panel shows the same data on a 
linear scale, normalized by the bolometric flux, with points connected by a spline 
fit. The dependence of 
the bolometric correction on color is the smallest in the SDSS $z$-band 
($\sim$0.9 $\mu$m).}
\label{sedBR}
\end{figure}

The broad-band colors (SED) of a galaxy depend both on its type and redshift (K correction). 
The dependence of optical and infrared colors on redshift is illustrated
in Fig.~\ref{Kcorr3IR} for SDSS ``main'' galaxies listed in the 2MASS XSC
and IRAS FSC. We used these diagrams to select a narrow redshift range
with a sufficient number of galaxies to construct the median SEDs for various
subsamples (i.e., we use the median of each color to construct the overall SED). 
The median SEDs constructed with GALEX, SDSS, and 2MASS photometry for
two subsamples of galaxies separated by $u-r$ color following Strateva
et al. (2001) and with redshifts in the $0.03 \leq z \leq 0.05$
range (small enough that the colors are essentially rest-frame) are
shown in the top panel in Fig.~\ref{sedBR} (when constructing 
SEDs as a function of wavelength, we use Vega to AB conversion for 2MASS magnitudes
from Finlator et al. 2000).

Galaxies with blue $u-r$ have all other colors, in the plotted wavelength range, bluer 
than galaxies with red $u-r$. Equivalently, the galaxy SEDs constructed 
with GALEX, SDSS, and 2MASS data are a nearly one parameter family (GALEX far-UV 
measurements do provide some additional information which cannot be extracted from 
SDSS and 2MASS broad-band measurements, see Ag\"ueros et al. 2005). In particular, 
we demonstrated in Section~\ref{Kpredict} that the 2MASS $K$-band flux can be 
predicted within 0.2 mag using SDSS $u$ and $r$ fluxes. Smol\v{c}i\'{c} et al.
(2006) discuss an even tighter one-dimensional behavior of galaxies at
wavelengths probed by SDSS. 

The bottom panel in Fig.~\ref{sedBR} shows the same SEDs as in the top panel,
except that a linear scale is used instead of a logarithmic scale, and the SEDs are
normalized by the bolometric flux. The bolometric flux is determined by 
integrating a spline fit to the 9 data points provided by GALEX, SDSS, and 
2MASS, and using Rayleigh-Jeans extrapolation at wavelengths longer than 2.2
$\mu$m. The data values shown in the figure are also listed in Table 2. 

The two normalized SEDs cross around the SDSS $z$
band ($\sim$0.9 $\mu$m); showing that the dependence of the bolometric correction 
for galaxies on color in the SDSS photometric system is the smallest in
the $z$-band. Hence, the flux measured
in this band provides the best approximation to the bolometric flux (up to a
constant; the important feature is the absence of color dependence). This band
also has the smallest K correction for the redshifts probed by the SDSS ``main''
galaxy sample (since the SED slope is the smallest around this wavelength
range), and is less sensitive to dust extinction than other SDSS bands. 
Thus, a good color-independent estimate of the bolometric flux (in the 0.2--2.2 
$\mu$m wavelength range) can be simply obtained from the expression
\begin{equation}
      (\nu \, F_\nu)_z = 0.58 \, F_{\rm bol},
\end{equation}
or, equivalently, 
\begin{equation}
       L_{\rm bol} = 20.2 \, 10^{-0.4\,M_z} L_\odot,
\end{equation}
where $M_z$ is the absolute SDSS $z$-band magnitude. The uncertainty
of these estimates is of order 5-10\% (including calibration
and SED integration errors, but not the individual $z$-magnitude
measurement error, which can exceed 10\% for faint galaxies).

We caution that the $z$-band should be used as a proxy for bolometric
flux only for galaxies at redshifts $\la 0.2$. For galaxies with larger
redshifts the 2MASS $J$-band measurement should be used instead (and the $H$
band for galaxies with redshifts beyond 0.6, though the number of such
galaxies detected by 2MASS may be extremely small).

\begin{table}
  \centering
  \label{medianseds}
   \caption{The median UV-to-near-IR SEDs, normalized by the bolometric flux ($\nu F_\nu / F_{bol} \equiv
    \lambda F_\lambda / F_{bol}$), 
        for blue ($u-r<2.22$) and red ($u-r>2.22$) galaxies with redshifts in the range 
        $0.03 \leq z \leq 0.05$.}
   \begin{tabular}{@{}cccc@{}}
    \hline
       bandpass & $\lambda$($\mu$m)  &  blue   & red \\
    \hline
       f   & 0.15      &  0.08   & $<$0.01 \\
       n   & 0.22      &  0.07   & $<$0.02 \\
       u   & 0.35      &  0.24   &  0.07   \\
       g   & 0.48      &  0.42   &  0.27   \\
       r   & 0.63      &  0.53   &  0.45   \\
       i   & 0.77      &  0.59   &  0.53   \\
       z   & 0.93      &  0.58   &  0.58   \\
       J   & 1.25      &  0.55   &  0.57   \\
       H   & 1.65      &  0.49   &  0.53   \\
       K   & 2.17      &  0.34   &  0.38   \\
    \hline
 \end{tabular}
\end{table}

\subsection{Do star-forming and AGN galaxies have different UV, IR, and radio properties? }
\label{SagnVSsf}

In Section~\ref{Sbpt} we demonstrated that star-forming and AGN galaxies, 
classified using {\it only} emission-line strengths, have different broad-band
optical properties such as $u-r$ and concentration index. Here we extend
that analysis and compare their UV, IR, and radio properties.

\subsubsection{The UV-color difference between star-forming and AGN galaxies}
\label{UVsfagn}

The GALEX-SDSS sample discussed here is fairly small,
and we only determined the median
far-UV-to-near-UV colors\footnote{For GALEX detections we require
$n<21$ or $f<21$ and correct magnitudes for interstellar extinction
using $A_f=2.97A_r$ and $A_n=3.23A_r$, where $A_r$ is the $r$--band
extinction from the maps of Schlegel, Finkbeiner \& Davis (1998) 
distributed with SDSS data. These coefficients were evaluated using the standard interstellar 
extinction law from Cardelli, Clayton \& Mathis (1989; M. Seibert, priv.~comm.). 
The median $A_r$ for the three AIS fields is 0.12, with a root--mean--square scatter 
of 0.02 mag.}, $f-n$ ($f$ and $n$ are AB magnitudes measured in the far- and near-UV
GALEX bands), for the two classes. We find that star-forming galaxies are 
bluer (median $f-n$ is 0.1$\pm$0.1) than AGNs (median $f-n$ is 0.5$\pm$0.1).
This is similar to the difference observed for $u-r$ 
(i.e., star-forming galaxies are bluer than AGNs). However, the analysis 
by Ag\"{u}eros et al. (2005) suggests that the origin of the color differences
is different for the $u-r$ and $f-n$ colors, because the latter
are much more sensitive to the presence of starbursts and AGNs, while the 
$u-r$ color reflects the bolometrically dominant stellar population, as 
suggested by the analysis described in Section~\ref{Kpredict}.

\subsubsection{The IR-color difference between star-forming and AGN galaxies}

\begin{figure} 
\centering
\includegraphics[bb=72 84 540 708, width=\columnwidth]{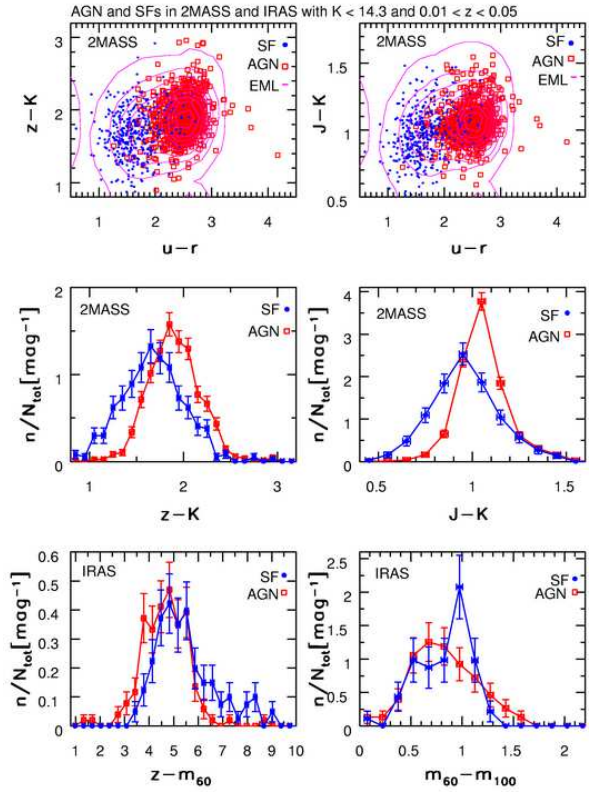}
\caption{The comparison of optical and infrared colors for galaxies 
detected by SDSS, 2MASS, and IRAS surveys, in a restricted redshift range 
($0.01 \leq z \leq 0.05$) to avoid K-correction
effects. Star-forming galaxies (designated by ``SF'') show 
more far-IR emission, relative to optical/IR, than AGN galaxies, but 
bluer optical and near-IR colors. }
\label{agnVSsf}
\end{figure}

\begin{figure} 
\centering
\includegraphics[bb=70 417 542 729, width=\columnwidth]{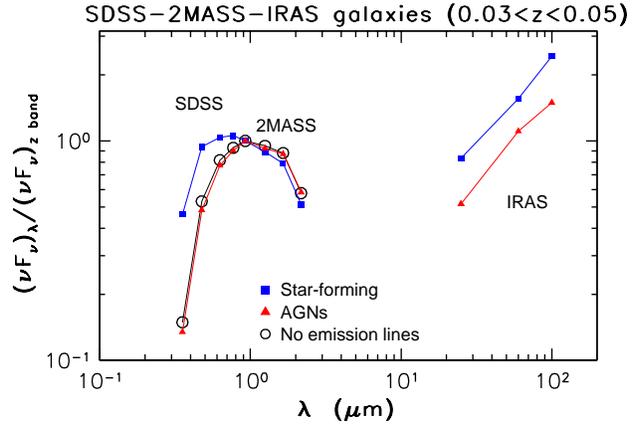}
\caption{The optical-to-far-IR SEDs, normalized to the SDSS $z$-band, shown
separately for galaxies without emission lines and AGN and star-forming 
emission-line galaxies detected by SDSS, 2MASS, and IRAS, and with redshifts 
in the $0.03 \leq z \leq 0.05$ range. Only the latter two subsamples have a sufficient
number of IRAS detections to characterize the far-IR SED.}
\label{sed}
\end{figure}

In order to study differences in infrared properties between star-forming
and AGN galaxies, we use the SDSS-2MASS and SDSS-IRAS samples, and restrict 
the redshift range to $0.01 < z < 0.05$ to avoid K-correction effects 
(further discussed below). The distribution of these galaxies in various 
optical/infrared color-color diagrams is shown in  
Fig.~\ref{agnVSsf}. In general, for any combination of optical and near-IR bands, 
star-forming galaxies are bluer than AGNs. For example, the $z-K$ and $J-K$
mean colors differ by 0.2 and 0.1 mag, respectively. According to the
Kolmogorov-Smirnov test (KS hereafter), these differences are highly statistically
significant, as is easily discernible from the two middle panels.
However, this relationship reverses when using far-IR bands; star-forming
galaxies have {\it redder} $z-m_{60}$ colors by $\sim$1 mag than AGNs
(the KS probability that the two subsamples are drawn from the same parent
sample is $\sim10^{-8}$) and have a much higher fraction of sources with
$z-m_{60} \ga 6$ (see the bottom left panel). This reversal is
illustrated in Fig.~\ref{sed} which compares the optical-to-far-IR 
SEDs, normalized to the SDSS $z$-band, for star-forming and AGN galaxies 
selected from a narrow redshift range. Since the $z$-band flux is a good
measure of the 0.2 - 2.2 $\mic$ bolometric luminosity, this implies that star-forming galaxies emit more IR 
radiation as a fraction of their 0.2--2.2 $\mic$ bolometric output, than do
AGN galaxies. This difference could be partially caused by a selection
effect; since the apparent $z$-band magnitudes of star-forming galaxies 
tend to be somewhat fainter that those of AGN galaxies due to differences in 
luminosity functions and sampled redshift range, they will be detected by IRAS
with the same probability only if they have somewhat redder $z-m_{60}$ colors.
However, this effect does not seem to be quantitatively sufficient to explain
the observed difference in median $z-m_{60}$ colors.

\begin{figure} 
\centering
\includegraphics[bb=80 197 582 779, width=\columnwidth]{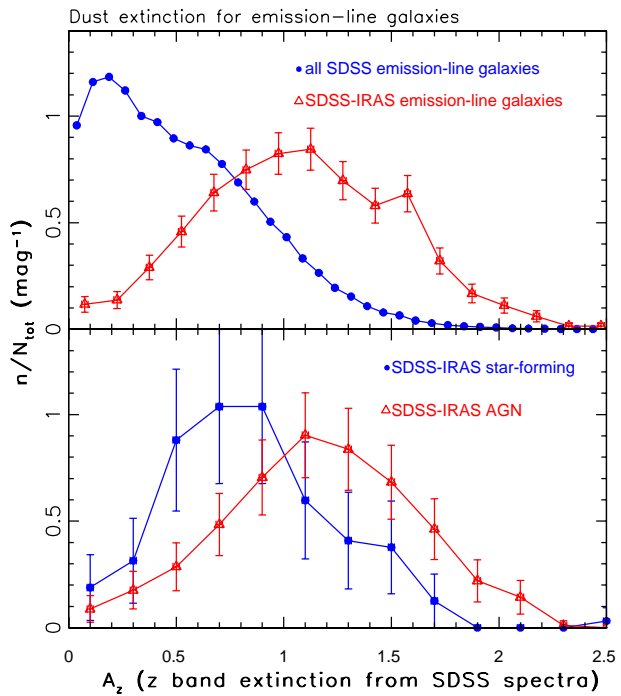}
\caption{The top panel compares the $z$-band dust 
extinction inferred from SDSS spectra by Kauffmann et al. (2003a), $A_z$, 
for all SDSS emission-line galaxies (dots) and for those also
detected by IRAS (triangles; analogous to the top panel in Fig.~\ref{IRASpred}, which also includes
galaxies without emission lines). The bottom panel compares the $A_z$
distributions for AGN (triangles) and star-forming galaxies (dots) detected by IRAS. Note 
that star-forming galaxies detected by IRAS typically have {\it smaller}
$A_z$ than AGN galaxies.}
\label{Az}
\end{figure}

There are several plausible explanations for the different median $z-m_{60}$ colors
for AGN and star-forming galaxies: star-forming galaxies 
could have warmer dust, significantly more dust, or more UV radiation
that is absorbed by dust and re-emitted in the far-IR\footnote{The observed difference
could also be due to different dust optical properties, i.e., different
ratio of far-IR to UV/optical opacity. While this possibility is not excluded
by our analysis, it is not necessary in order to explain the observed trends. Similarly,
we ignore the possible effects of dust geometry, e.g., the dust around AGN could
have significantly different distribution than interstellar dust (e.g., Nenkova,
Ivezi\'{c} \& Elitzur 2002).} than AGN galaxies (of course, these possibilities 
are not mutually exclusive). Different dust temperatures is probably
ruled out two types have very similar $m_{60}-m_{100}$ color distributions; this color 
is much more sensitive to dust temperature than to the amount of dust; for a detailed 
discussion see e.g., Ivezi\'{c} \& Elitzur 1997), as shown in the bottom right panel 
in Fig.~\ref{agnVSsf}. In order to test the hypothesis that the difference is caused 
by a different amount of dust, we compared the distributions of extinction estimates, 
$A_z$, discussed in Section~\ref{SAz}. As discernible from the bottom panel in 
Fig.~\ref{Az}, star-forming galaxies detected by IRAS typically have {\it smaller} 
$A_z$ than AGN galaxies, thus ruling out this hypothesis. 

Thus, it appears that star-forming galaxies emit 
more far-IR radiation (as a fraction of bolometric flux in the 0.2--2.2 $\mic$ 
wavelength range) than AGN galaxies, despite having smaller $A_z$ of dust at
similar temperatures, because they have significantly more UV radiation
(again, as a fraction of bolometric flux in the 0.2--2.2 $\mic$
wavelength range) that is processed into far-IR wavelength range.
This conclusion is supported by GALEX data discussed by Ag\"{u}eros et al.
(2005), who find a correlation between $f-n$ and contribution of 
the UV flux to the bolometric flux. Since 
star-forming galaxies tend to have bluer $f-n$ colors than AGNs (see 
Section~\ref{UVsfagn}), this implies that their UV flux contributes more
to the bolometric flux than for AGN galaxies. Indeed, once a larger
SDSS-GALEX-IRAS sample is available, the expected strong correlation between 
$f-m_{60}$ and $A_z$ (similar to, and perhaps stronger than, the correlation shown 
in Fig~\ref{IRASpred}) can be directly tested\footnote{The analysis of Buat et al. 
(2004) appears to support this expectation: they find that interstellar extinction
for a far-IR flux-limited sample is higher than for a UV flux-limited
sample. However, their estimates of interstellar extinction are derived 
from the UV-IR colors (rather than from independent data), and thus are 
strongly correlated with the sample flux limits. That is, the same
effect would be obtained even if there were no correlation between 
the amount of dust in a galaxy and the UV-IR colors, and hence their
analysis {\it cannot} be used to support our interpretation.}. 

In a recent paper based on similar data, Pasquali, Kauffmann \& Heckman (2005,
hereafter PKH) found that the ``AGN exhibit a significant excess in far-IR 
emission relative to star-forming galaxies'', which apparently contradicts the 
results presented here. However, it should be pointed out that here we discuss 
the amount of IR radiation (at 60 $\mic$) as a fraction of bolometric output 
in the 0.2--2.2 $\mic$ range (i.e., $z-m_{60}$ color), while PKH discuss the 
IR luminosity {\it per unit stellar mass}. Furthermore, their statement is 
valid for carefully selected pairs of AGN and star-forming galaxies that
have similar physical characteristics such as stellar mass, color, size, etc.
(it is also noteworthy that the distributions of stellar mass and stellar 
mass-to-light ratio are very different for AGN and star-forming galaxies,
as shown by Kauffmann et al. 2003a). Hence, these are two independent,
rather than contradictory, findings.

\subsubsection{The radio-IR correlation}
\label{radioIR}

\begin{figure} 
\centering
\includegraphics[bb=72 474 540 708, width=\columnwidth]{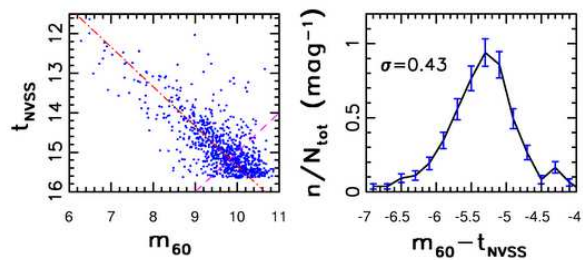}
\caption{The left panel shows the radio 20 cm magnitude measured by NVSS
as a function of 60 $\mu$ flux measured by IRAS, for 948 SDSS galaxies 
with a counterpart in IRAS FSC within 30 arcsec and in the NVSS catalog 
within 15 arcsec. The right panel shows the distribution of far-IR-radio
color, $m_{60}-t_{NVSS}$, for 568 galaxies with $m_{60}+t_{NVSS}<25$ 
(upper left region, away from the dashed line, shown in the left panel). 
The median value of $m_{60}-t_{NVSS}$ (-5.31) is shown as the diagonal 
dot-dashed line in the left panel, and the rms distribution width determined 
from the interquartile range is marked in the right panel.}
\label{radioPlot2}
\end{figure}

The large number of SDSS-IRAS-NVSS emission-line galaxies allows us to examine 
whether the well-known narrow distribution of the far-IR-to-radio flux ratio 
(van der Kruit 1971) is the same for (optically classified) star-forming and 
AGN galaxies. This correlation was
interpreted by Helou et al. (1985) as a consequence of coupling between 
infrared thermal dust emission and radio non-thermal synchrotron emission,
and it is not known whether the details of this coupling are the same
for star-forming and AGN galaxies (e.g., see Bell 2003). In addition, both 
far-IR and radio emission are used as probes for ongoing and recent star 
formation (Hopkins et al. 2003 and references therein), and thus 
a robust measurement of this correlation is important for the comparison
of studies based on different data sets.

\begin{figure} 
\centering
\includegraphics[bb=85 77 516 470, width=\columnwidth]{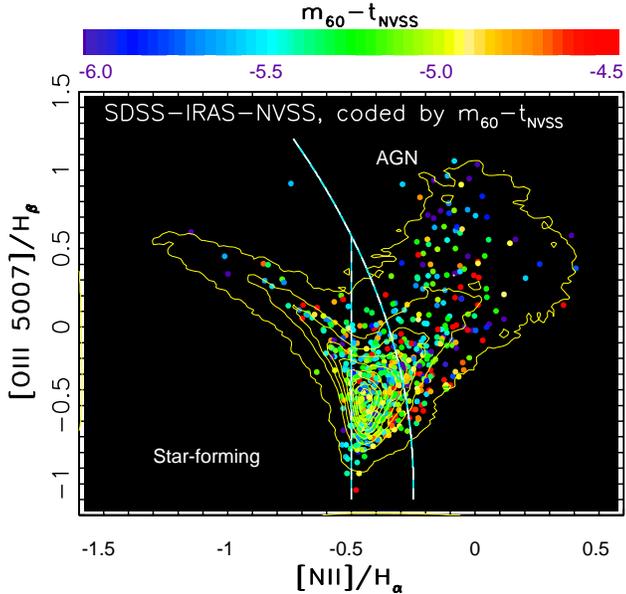}
\caption{The distributions of SDSS-IRAS-NVSS emission-line galaxies (dots) and 
all SDSS emission-line galaxies (contours) in the BPT diagram. The dots are colored
according to their $m_{60}-t_{NVSS}$ color, as shown in the legend on top.
{\it Both} AGN and star-forming galaxies are found in this sample, and they 
all follow the well-known radio-IR correlation.}
\label{BPTcondon}
\end{figure}

The left panel in Fig.~\ref{radioPlot2} shows the radio 20 cm magnitude measured 
by the NVSS as a function of 60 $\mu$m magnitude measured by IRAS for 948 SDSS ``main''
galaxies with a counterpart in IRAS FSC within 30 arcsec, and in the NVSS catalog 
within 15 arcsec. The strong correlation between the two fluxes\footnote{Sometimes
this correlation is shown in the luminosity vs. luminosity form, which 
boosts the impression of correlation strength. However, this is not a good
practice since even uncorrelated measurements could give such an impression
if the redshift distribution is sufficiently broad.} is evident. Only about 1\% of sources 
have anomalously bright radio emission (by about $\sim$2 mag), a fraction that 
is consistent with contamination by random associations (the same fractions of AGN
and star-forming galaxies are found among those 1\% of sources, as in the whole
sample; they are all bright point sources at 20 cm), but could also be due to 
radio-loud objects. In order to avoid the effect 
of faint flux limits on the distribution of the far-IR-to-radio flux ratio 
(i.e., the $m_{60}-t_{NVSS}$ color),
we restrict the sample to 568 galaxies with $m_{60}+t_{NVSS}<25$ (a condition 
perpendicular to the $m_{60}-t_{NVSS} = const.$ lines in the $t_{NVSS}$ vs. $m_{60}$
plane, see the dashed line in the left panel in Fig.~\ref{radioPlot2}), and show the distribution of 
the $m_{60}-t_{NVSS}$ color in the right panel. The median value of the $m_{60}-t_{NVSS}$ 
color is -5.31$\pm$0.02 (statistical errors only), with an equivalent Gaussian 
width (determined from interquartile range) of 0.43 mag. Such a narrow width 
is quite remarkable given so different wavelengths and survey technologies,
and represents a strong constraint for the theories of coupled radiation
mechanisms (e.g., Helou et al. 1985). It also places an upper limit of
0.4 mag on the IRAS photometric error for galaxies.

A very similar analysis of IR-radio correlation based on a smaller sample (176 
UGC galaxies) was presented by Condon \& Broderick (1988), who used a slightly 
different parameter $u=-0.4(m_{60}-t_{NVSS})$ (sometimes also called the
$q$ parameter, e.g., Bell 2003). They find a peaked distribution 
similar to that shown in Fig.~\ref{radioPlot2}, with 63 galaxies in the peak. 
The position of that peak corresponds to $m_{60}-t_{NVSS} = -5.05 \pm 0.05$, 
in good agreement with our analysis based on a $\sim$10 times larger sample. 
The width of the $m_{60}-t_{NVSS}$ distribution determined here (0.43 mag) corresponds
to a width of 0.17 for the $u$ (or $q$) distribution, somewhat smaller than $\sim$0.26
obtained in previous studies (e.g., Yun, Reddy \& Condon 2001, Bell 2003). 
Condon \& Broderick proposed that galaxies in this peak ($m_{60}-t_{NVSS} < -4$ ) 
are dominated by starbursts, while AGNs (``monsters'' in their terminology) have 
too weak IR emission to be detected by IRAS, implying $m_{60}-t_{NVSS} > -4$. 

The high-quality SDSS spectra and corresponding BPT-diagram based separation of 
emission-line galaxies into AGN and star-forming galaxies allows us to examine 
IR-radio correlation in detail. We find that over 1/3 of SDSS-NVSS-IRAS galaxies 
with emission lines can be reliably classified as an AGN based on their position 
in the BPT diagram, with an additional 50\% in the transition region (``unknown''
sources, using classification described in Section~\ref{Sbpt}). That is, in addition to 
star-forming galaxies, optically-classified AGN galaxies also follow the tight 
IR-radio correlation. Hence, not all the ``monsters'' are confined to 
$m_{60}-t_{NVSS}>-4$ (assuming that they are correctly recognized by BPT analysis).

We note that the AGN/star-forming galaxy separation in the BPT diagram adopted here is 
fairly conservative (see Section~\ref{Sbpt}). Figure~\ref{BPTcondon} demonstrates
that  the majority of SDSS-IRAS-NVSS emission-line galaxies have ambiguous 
classification in the BPT diagram. However, it is easily discernible that a 
substantial fraction of these galaxies have optical emission-line strength
ratios fully consistent with an AGN classification. The figure also shows that 
the position of a galaxy in the BPT diagram and the $m_{60}-t_{NVSS}$ color do 
not appear to be correlated. Visual inspection of SDSS $g, r, i$ color composite 
images convincingly shows that {\it both} AGN and star-forming subsamples are 
dominated by galaxies with spiral morphology, as already pointed out by Condon 
\& Broderick (1988). About half of them show disturbed morphology and nearby 
companions.
\begin{figure} 
\centering
\includegraphics[bb=20 527 302 759, width=\columnwidth]{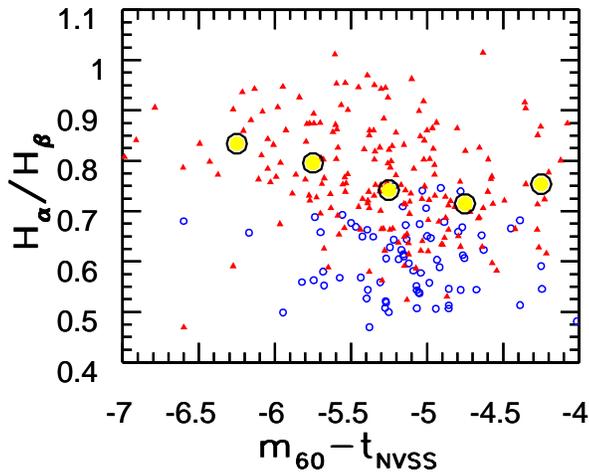}
\caption{The distribution of AGN and star-forming galaxies in the radio-IR color, 
$m_{60}-t_{NVSS}$, vs. optical emission-line strength ratio, $H_\alpha/H_\beta$.
Small blue circles show star-forming galaxies and red triangles AGN galaxies. 
The large circles are the median values of $H_\alpha/H_\beta$ in narrow
$m_{60}-t_{NVSS}$ bins for AGN galaxies, and demonstrate that the strength
of IR emission for AGN galaxies, relative to their radio emission, increases 
with the $H_\alpha/H_\beta$ line strength ratio.}
\label{HaHb_IRradio}
\end{figure}

\begin{figure} 
\centering
\includegraphics[bb=85 77 516 490, width=\columnwidth]{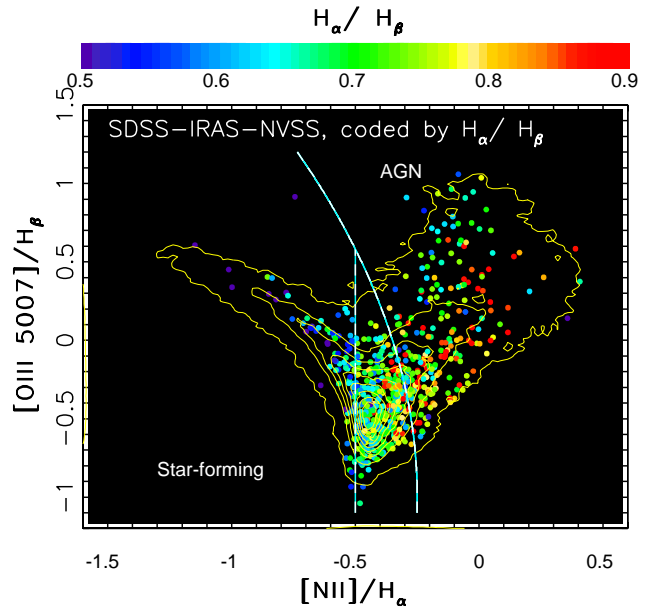}
\caption{The distributions of SDSS-IRAS-NVSS emission-line galaxies (dots) and 
all SDSS emission-line galaxies (contours) in the BPT diagram. The dots are colored
according to their $H_\alpha/H_\beta$ line strength ratio, as shown in the legend 
on top. Note that objects with the largest values of  $H_\alpha/H_\beta$ ratio (red)
are found almost exclusively in AGN region.}
\label{BPTcondon2}
\end{figure}

Using classification based on the BPT diagram, we compute and compare the 
slope of IR-radio correlation separately for AGN and star-forming galaxies.
Two subsamples with 128 AGN and 46 star-forming galaxies yield median $m_{60}-t_{NVSS}$ 
values of $-5.35\pm0.05$ and $-5.14\pm0.05$, and widths of 0.61 and 0.38 mag, 
respectively\footnote{The $m_{60}-t_{NVSS}$ 
color measurement should not be interpreted as implying a power-law 
spectral energy distribution between 60 $\mu$m and 20 cm with the power-law 
index of $\sim$0.6 ($F_\nu \propto \nu^\alpha$). For example, galaxies have 
$\alpha \sim -0.5$ between 6 and 20 cm (Ivezi\'{c} et al. 2004a), and 
$m_{60}-m_{100}\sim1$, implying $\alpha \ga 1$ between 100 $\mu$m and 6 cm, 
that is, a steeper decrease of flux with wavelength, than implied by the 
$m_{60}-t_{NVSS}$ colors.}. The difference between the 
medians is $\sim 3\sigma$ significant, and suggests that the details 
of radiation coupling mechanisms may be different for star-forming
and AGN galaxies (i.e., among galaxies that follow IR-radio correlation,
star-forming galaxies appear to show $\sim$20\% more radio emission, relative 
to far-IR, than AGN galaxies). We note that this 
result should be considered somewhat tentative because there may be systematic 
effects that are not included when estimating uncertainties in the 
medians\footnote{Some of this difference could be due to systematically
different radio morphology, an interesting possibility that is beyond the scope
of this work. It is noteworthy that we did not find a correlation between the 
$m_{60}-t_{NVSS}$ color and the radio spectral slope between 20 cm and 92 cm.}, 
and thus its significance could be overestimated. It is also noteworthy
that none of star-forming galaxies have $m_{60}-t_{NVSS}>-4$, while this is
true for 4\% of AGNs (that is, an opposite trend than for medians).

We searched for possible correlations between $m_{60}-t_{NVSS}$ and other 
observables that could perhaps explain the different median values of this
color for AGN and star-forming subsamples (either as selection effects, bad data, 
bad analysis method, or astrophysics). We analyzed quantities such as colors, 
redshift, stellar mass, luminosity, angular size, dust extinction estimate $A_z$, 
etc., and the only quantity that appears to have an effect on $m_{60}-t_{NVSS}$ is
the $H_\alpha/H_\beta$ ratio, as shown in Fig.~\ref{HaHb_IRradio} (see
also Fig.~\ref{BPTcondon2} which illustrates correlation between the 
$H_\alpha/H_\beta$ ratio and the position in BPT diagram). Star-forming 
galaxies are confined to the $H_\alpha/H_\beta < 0.7$ region, while AGN galaxies
span the whole observed range of $H_\alpha/H_\beta$. Furthermore, as the 
$H_\alpha/H_\beta$ line strength ratio increases, the strength of IR emission
for AGN galaxies, relative to radio emission, also increases (i.e., the $m_{60}-t_{NVSS}$
color becomes bluer). Since the measurement of $m_{60}-t_{NVSS}$ is fully
independent of the $H_\alpha/H_\beta$ measurement, this behavior provides additional
support for the 3$\sigma$ significant difference in the slopes of IR-radio correlation
for star-forming and AGN galaxies (note, however, that the AGN presence 
could have an effect on the measured $H_\alpha/H_\beta$ ratio). It is noteworthy that 
there is no evidence for a similar behavior in the $m_{60}-t_{NVSS}$ vs $A_z$ diagram,
although some degree of correlation exists between $H_\alpha/H_\beta$ and $A_z$.
This may mean that the effects of gas and dust on infrared and radio emission
are more complex than implied by a simple linear IR-radio correlation (for
a detailed discussion of this possibility see Bell 2003). 

Given that both AGN and star-forming subsamples of SDSS-NVSS-IRAS
galaxies follow very similar, if not identical, radio-IR correlation, it is 
interesting to investigate what fraction of the full SDSS-NVSS and 
SDSS-IRAS subsamples could follow this correlation (that is, galaxies 
that are detected by only two, instead of all three, surveys).
We perform this analysis by using the observed radio-IR correlation to 
predict $t_{NVSS}$ for SDSS-IRAS galaxies, or $m_{60}$ for SDSS-NVSS galaxies. 
In the second step we select galaxies with predicted fluxes 1 mag brighter 
than the faint limit of the corresponding third catalog (to account for the
scatter due to photometric errors), and then determine what fraction of these galaxies 
are actually detected. For example, if {\it every} IRAS galaxy follows 
IR-radio correlation, then we estimate that $>90\%$ of SDSS-IRAS galaxies 
with predicted $t_{NVSS}<15.5$ should be detected by NVSS. We find that 
indeed 92\% of SDSS-IRAS galaxies with $t_{NVSS}^{predicted} =  m_{60} + 5.3 < 14.5$ 
are detected by NVSS. 

The converse is not true. Only about 1/3 of SDSS-NVSS {\it emission-line} 
galaxies with $m_{60}^{predicted} = t_{NVSS} - 5.3 < 9$ are detected
by IRAS. This is consistent with a hypothesis that there is another 
source of radio emission, in addition to the component that is correlated
with IR emission. The elevated radio emission is then responsible 
for the NVSS detection, but IR emission is too weak for an IRAS
detection. Presumably, those AGNs that are detected by both NVSS and
IRAS are mostly radio-quiet, while those with elevated radio emission
are mostly radio-loud. Not surprisingly, SDSS-NVSS emission-line
galaxies without IRAS detection have typically redder $u-r$ colors than
those detected by both NVSS and IRAS (the difference  in medians is 0.6
mag). The detection fraction by IRAS is different for AGN and
star-forming subsamples of SDSS-NVSS sample: while 50\% of star-forming
SDSS-NVSS galaxies are detected by IRAS, this is true for only 20\% of
AGNs (for galaxies with $m_{60}^{predicted} < 9$). These statistics
indicate that the {\it majority} (80\%) of emission-line galaxies 
classified {\it optically} as AGNs emit more flux at radio wavelengths than 
implied by their IR fluxes and the mean IR-radio correlation. Again, 
practically all of those remaining 20\% AGNs from SDSS-NVSS subsample 
that are also detected by IRAS do follow a very tight IR-radio correlation.

\subsection{ Adding panchromatic information to the BPT diagram }
\label{multiBPT}

\begin{figure} 
\centering
\includegraphics[bb=72 190 540 602, width=\columnwidth]{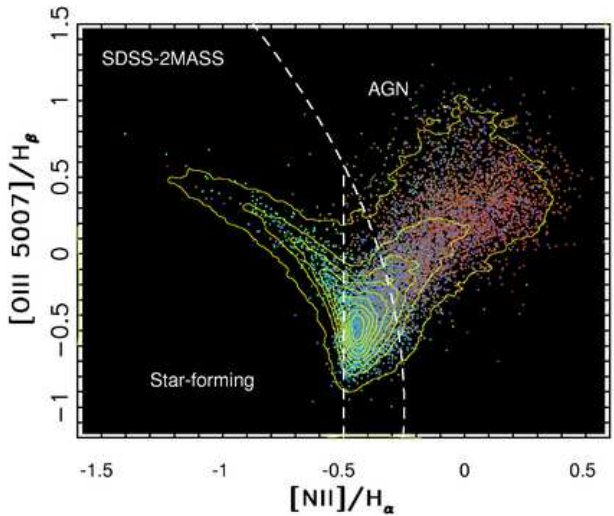}
\caption{The comparison of the distributions of SDSS-2MASS emission-line
  galaxies (dots) and all SDSS emission-line galaxies (contours) in the BPT diagram. The dots are colored
according to their position in the concentration index vs. $u-r$ diagram, shown
in the top right panel in Fig.~\ref{BPTurC}. The AGN-to-SF number ratio for SDSS-2MASS 
galaxies is 10.1, while it is 1.6 for the whole SDSS sample.}
\label{BPT2MASS}
\end{figure}

\begin{figure} 
\centering
\includegraphics[bb=72 190 540 602, width=\columnwidth]{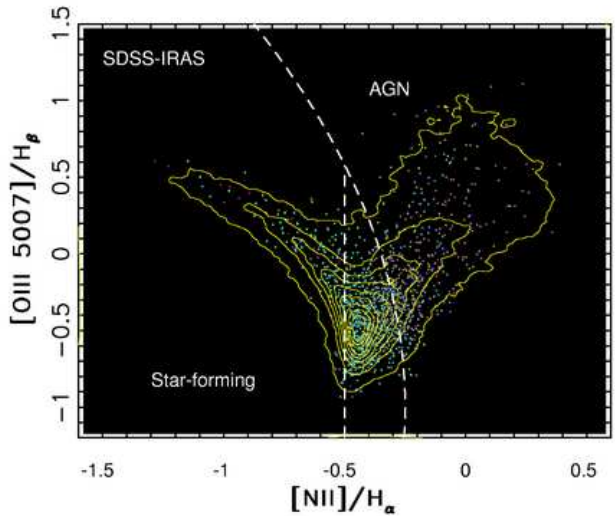}
\caption{Analogous to Fig.~\ref{BPT2MASS}, except for SDSS-IRAS galaxies.
The AGN-to-SF number ratio for SDSS-IRAS galaxies is 2.7.}
\label{BPTIRAS}
\end{figure}

\begin{figure} 
\centering
\includegraphics[bb=72 190 540 602, width=\columnwidth]{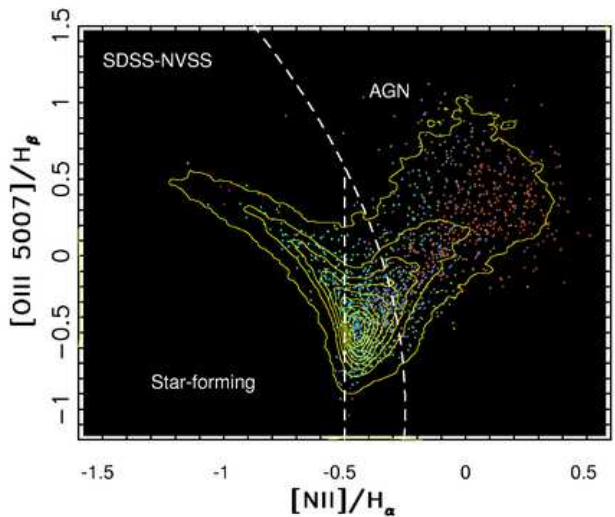}
\caption{Analogous to Fig.~\ref{BPT2MASS}, except for SDSS-NVSS galaxies.
The AGN-to-SF number ratio for SDSS-NVSS galaxies is 4.7.}
\label{BPTNVSS}
\end{figure}

\begin{figure} 
\centering
\includegraphics[bb=80 50 522 439, width=\columnwidth]{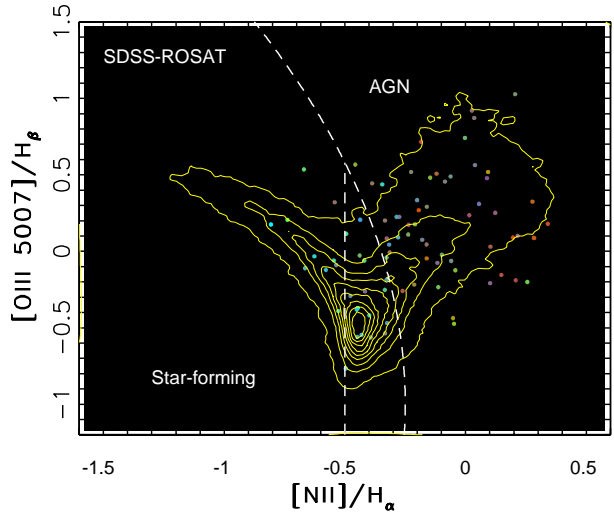}
\caption{Analogous to Fig.~\ref{BPT2MASS}, except for SDSS-ROSAT galaxies.
The AGN-to-SF number ratio for SDSS-ROSAT galaxies is 5.7.}
\label{BPTROSAT}
\end{figure}

\begin{figure} 
\centering
\includegraphics[bb=20 50 560 760, width=\columnwidth]{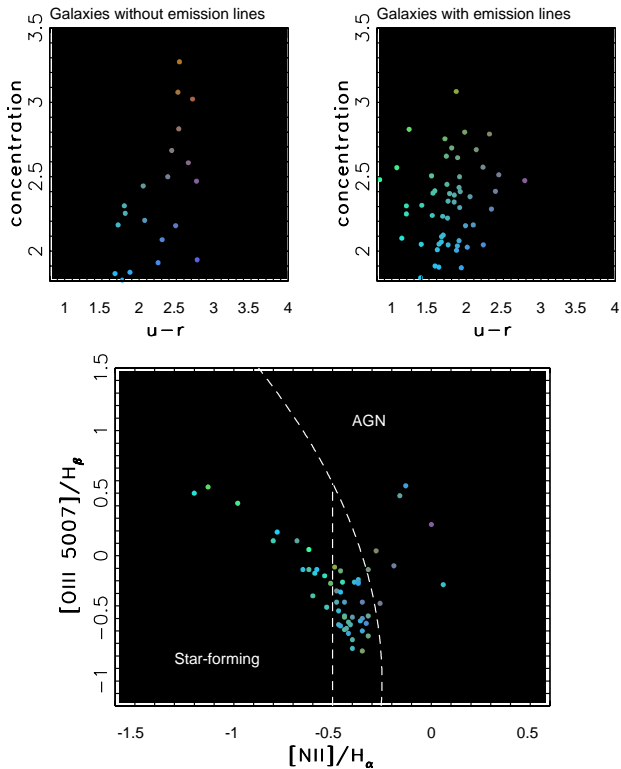}
\caption{Analogous to Fig.~\ref{BPTurC}, except for SDSS-GALEX galaxies.
The lower limit on AGN-to-SF number ratio for SDSS-GALEX galaxies is at least 0.1, and
it could be as high as 0.3.}
\label{BPT2GALEX}
\end{figure}

\begin{figure*}
\centering
\includegraphics[bb=68 72 543 720, angle=270, width=1.9\columnwidth]{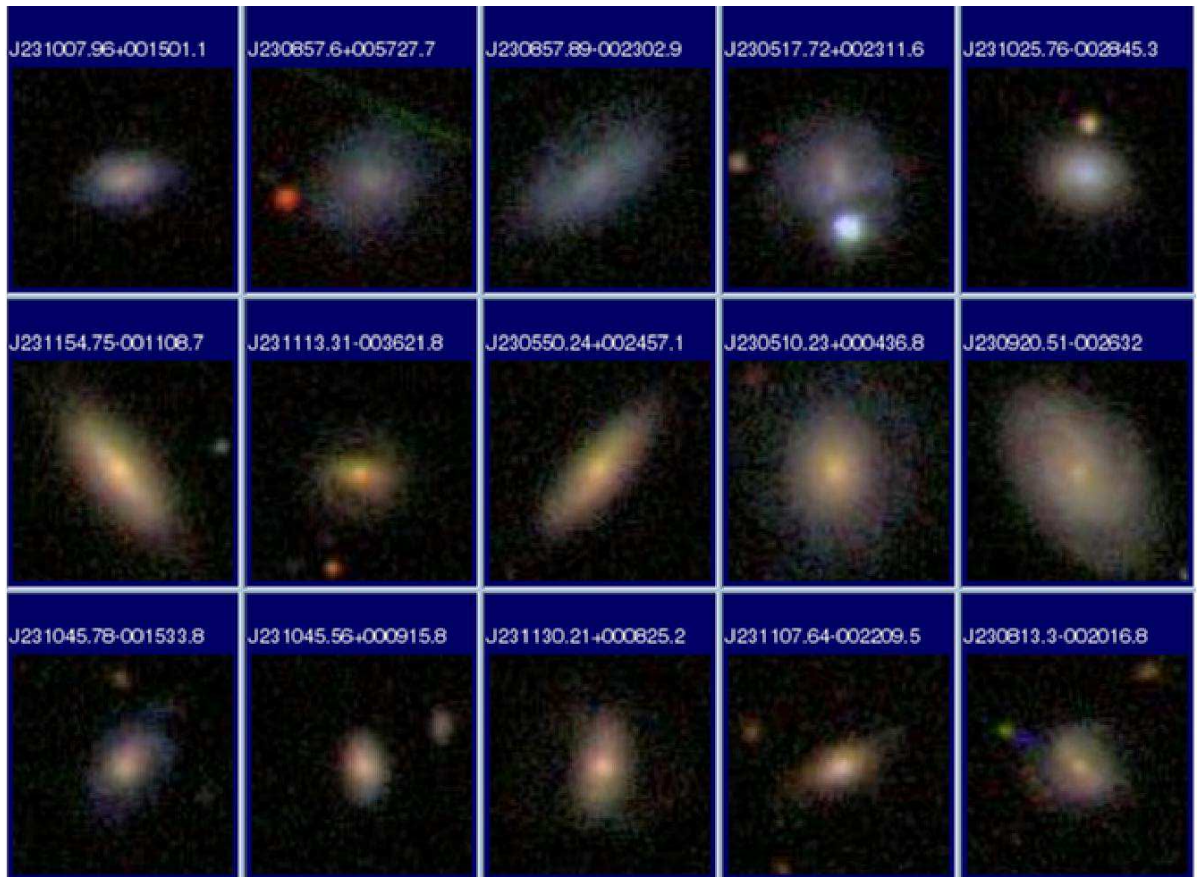}
\caption{{\it g, r, i} composite SDSS images of SDSS-GALEX galaxies
  randomly chosen from three subsamples classified using emission-line strengths measured
 from SDSS spectra (Kauffmann et al. 2003a). The first row shows 
images of star--forming galaxies, the second of AGN; and galaxies in the third row 
have uncertain classifications. North is up, and the images are roughly $25\arcsec$ 
on a side.}
\label{GALEXmorph}
\end{figure*}

In preceding Sections we have discussed the broad-band panchromatic properties
of galaxy samples selected by their emission-line properties with the aid
of the BPT diagram (see Section~\ref{Sbpt}). In this Section we study how the 
morphology of galaxy distribution in the BPT diagram changes when requiring
detections at other wavelengths (see Figs.\ref{BPT2MASS} to \ref{BPT2GALEX}).

The AGN-to-star-forming galaxy number ratio is 1.6 for the whole SDSS sample.
Although we exclude a large number of galaxies with uncertain classification 
(see Section~\ref{Sbpt}), this ratio is a good relative measure of the changes 
in the BPT diagram\footnote{Note that the AGN-to-star-forming galaxy
  number ratio depends on the adopted cutoff for the emission line
  detection significance. Higher values than the 3$\sigma$ adopted here
  (see Section~\ref{Sbpt}) would result in a lower AGN-to-star-forming
  galaxy number ratio because many AGNs are very weak-lined LINERs
  (Heckman et al. 2004).}. We find that the AGN-to-star-forming galaxy
number ratio is systematically larger for subsamples with detection at
other wavelengths, except for SDSS-GALEX sample, and it is the largest
for SDSS-2MASS sample (10.1). Such a high ratio for SDSS-2MASS sample is
a consequence of fairly bright K-band flux limit, and the fact that AGN
galaxies have redder optical-to-near-IR SEDs than star-forming galaxies.

The AGN-to-star-forming galaxy number ratio is the smallest for SDSS-GALEX sample
(0.1, but it could be as high as 0.3 if the unclassified galaxies are 
dominated by AGN galaxies, see Fig.~\ref{BPT2GALEX}). The lower limit on this ratio 
is sufficiently high to exclude the possibility that SDSS-GALEX galaxies represent a clean
sample of starburst galaxies. In order to present further evidence for this 
claim, we have visually inspected SDSS $g, r, i$ color composite images of 
these galaxies (a total of 55) and found that the classification based on 
emission-line strengths is well correlated with morphology. 
SDSS images of random subsamples of AGN, star-forming, and unclassified
galaxies are shown in Fig.~\ref{GALEXmorph}. Clear morphological differences between 
galaxies classified as star-forming and as AGN are easily discernible, with 
the latter being more centrally concentrated. This further demonstrates that 
{\it at least some GALEX/SDSS galaxies are more likely to be AGN than star-forming}.

\subsection{An improvement of the K band flux prediction}
\label{KpredImprove}

In Section~\ref{Kpredict} we showed that it is possible to estimate the 
$K$-band magnitude with a scatter as small as $\sim$0.2 mag using only
SDSS data. In this Section we explore whether the residuals between
predicted and 2MASS $K$-band magnitudes correlate with several model-dependent
quantities determined by Kauffmann et al. (2003a), and whether the residuals
show the same behavior when AGN and star-forming galaxies are treated separately.

\begin{figure} 
\centering
\includegraphics[bb=147 72 435 720, width=\columnwidth]{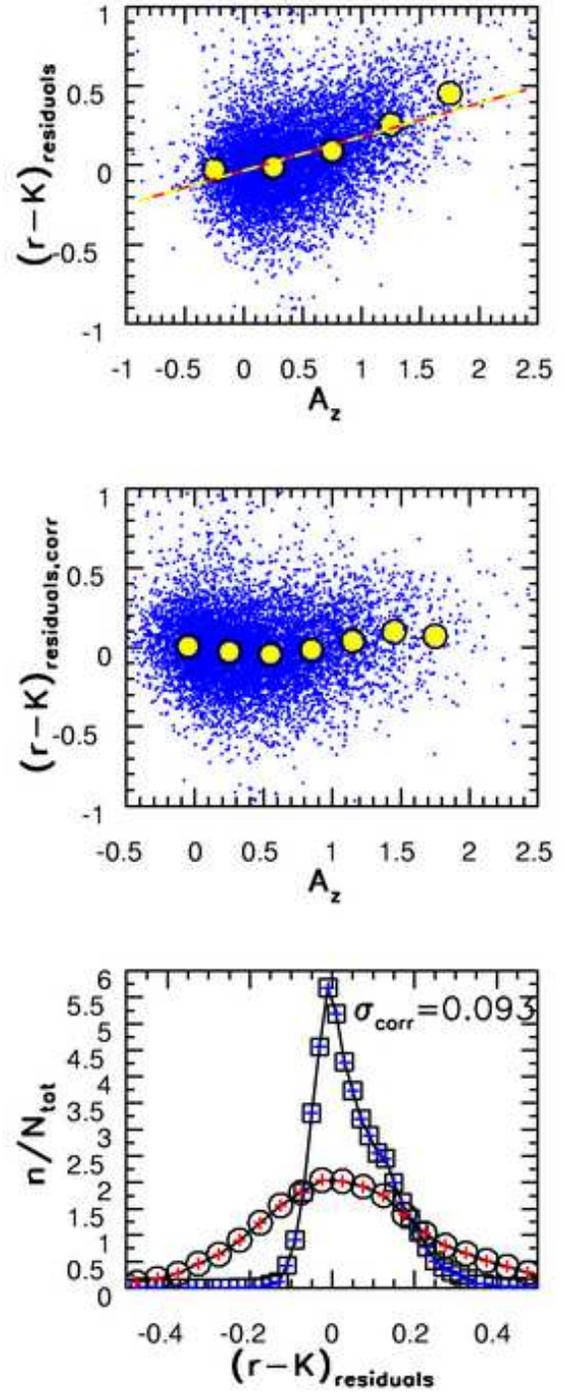}
\caption{The top panel shows the dependence of residuals between predicted and 
measured K-band magnitudes on the $z$-band dust extinction, $A_z$, inferred from 
SDSS spectra by Kauffmann et al. (2003a). Individual galaxies are shown by small
symbols, and large circles show the median values in $A_z$ bins. The best-fit 
straight line to these medians is also shown. The residuals corrected for this 
median trend are shown in the middle panel. The bottom panel compares the 
distributions of uncorrected residuals (circles) and the corrected ones (squares). 
The distribution width for the latter is only about one half of that for the former.}
\label{Kpred2}
\end{figure}

\begin{figure} 
\centering
\includegraphics[bb=72 182 540 610, width=\columnwidth]{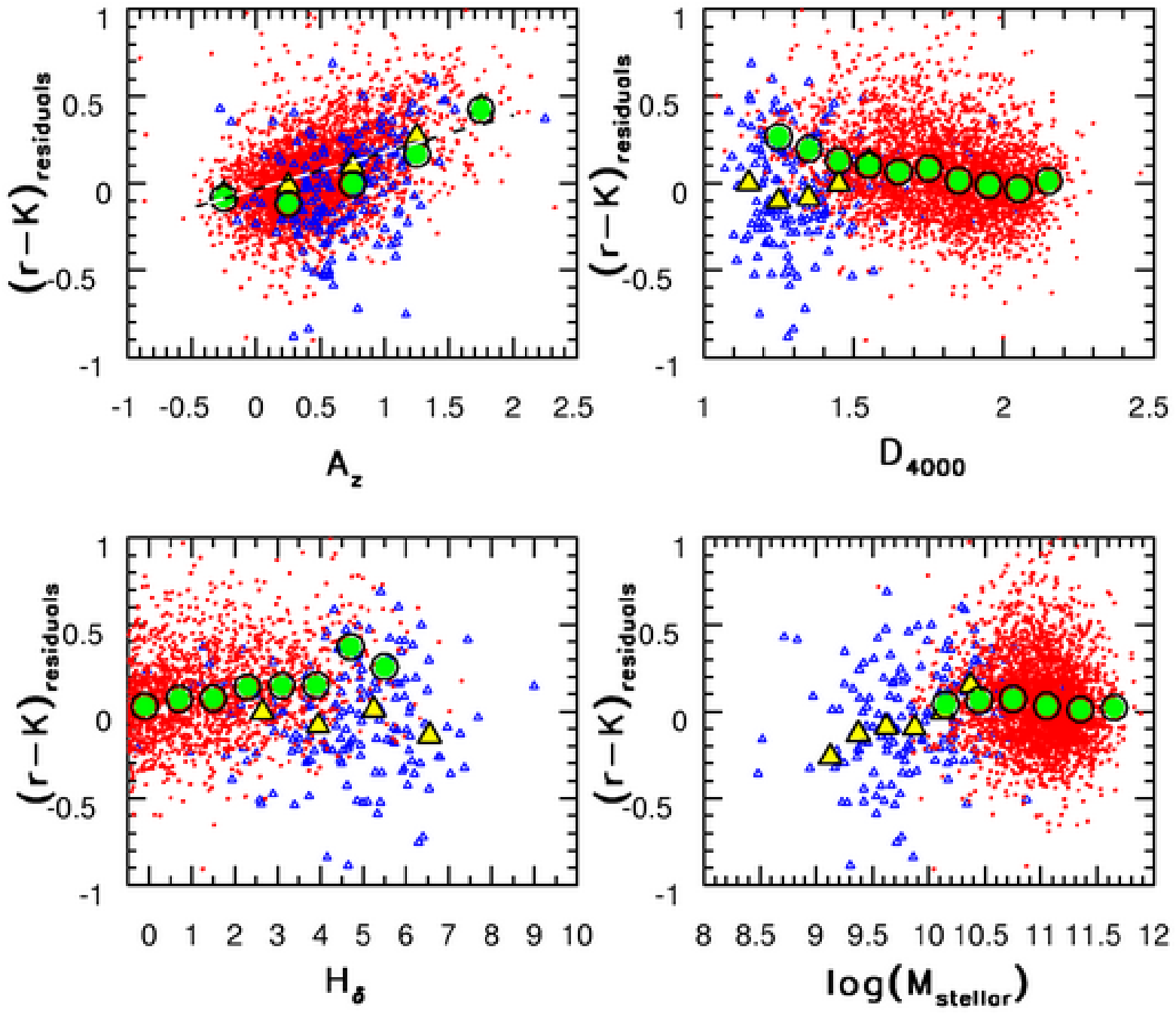}
\caption{The top left panel is analogous to the top panel in Fig.~\ref{Kpred2},
except that here AGN (dots and circles) and star-forming galaxies (small and 
large triangles) are treated separately. The straight line is the same line 
as in the top panel in Fig.~\ref{Kpred2}. The other three panels are analogous
to the top left panel, except that they show the dependence of residuals  on 
the 4000 \AA\ break ($D_{4000}$), the strength of $H_\delta$ line, and stellar
mass.}
\label{Kpred3}
\end{figure}

The strongest correlation between the $K$-band SDSS-2MASS residuals and another 
quantity is found for $A_z$, the galaxy dust content discussed in detail in 
Section~\ref{SAz}. The top panel in Fig.~\ref{Kpred3} illustrates this correlation.
A best straight line fit is given by 
\begin{equation} 
    \Delta(r-K)^\ast = 0.213~A_z - 0.033
\end{equation}
This correlation may be interpreted as the effect of dust on the observed
$r$-band flux (however, note that the simple extinction screen approximation
is probably not appropriate). When this correlation is subtracted from $(r-K)^\ast$ 
given by the eq.~\ref{Kpredeq}, the width of the residuals distribution decreases 
by a factor of 2, to 0.1 mag! In other words, given the $u-r$ color, redshift, and 
$A_z$ determined using SDSS data, the 2MASS $K$-band measurements can be predicted with
a scatter of only 0.1 mag. Most of this scatter can be attributed to the 
measurement errors. For example, assuming conservative lower limits for 
errors in $u-r$ (0.03 mag), $K$ (0.03 mag, Jarrett et al. 2000), and $R_{50}^z$ (2\%),
the expected scatter due to measurement errors is 0.09 mag. Hence, the observed
residual scatter of 0.1 mag is likely dominated by measurement errors.
We note that the final distribution of the $r-K$ residuals shown in the 
bottom panel in Fig.~\ref{Kpred2} is skewed. This could be due to the fact that 
the dependence of the $r-K$ residuals on $A_z$ was fit by a straight line, while 
the data display some curvature. Also, it could be that there are two subpopulations 
of galaxies that have slightly different SEDs. 

Since the residual astrophysical scatter is apparently much smaller than 0.1 mag,
one is tempted to conclude that 2MASS measurements are not required to study 
SDSS galaxies. This would not be a valid conclusion for at least two reasons. 
First, the measurement error for 2MASS $K$ band magnitudes is $\sim$0.03 mag,
which is smaller than the residual scatter for predicted $K$ band magnitudes
(0.1 mag). Second, we have not investigated morphological properties of galaxies
in the near-IR, where smaller dust extinction could reveal features not visible
in the optical wavelength range.

We have also studied the correlation between the $K$-band SDSS-2MASS residuals and 
the 4000 \AA\ break ($D_{4000}$), the strength of $H_\delta$ line, and stellar
mass (Fig.~\ref{Kpred3}). None of the correlations is as strong as the correlation
with $A_z$. As shown in Fig.~\ref{Kpred3}, the overall behavior of AGN and
star-forming subsamples are similar to each other, though not identical.

\subsubsection{ Comparison with Bruzual \& Charlot models }

\begin{figure} 
\centering
\includegraphics[bb=20 50 560 520, width=\columnwidth]{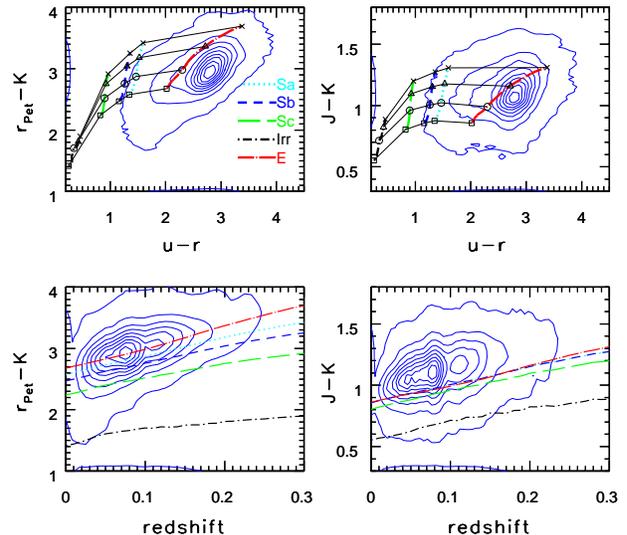}
\caption{A comparison of the observed galaxy distribution (contours) in representative
color-color and color-redshift diagrams and the Bruzual \& Charlot (1993,
2003) model predictions. The SEDs for five representative
synthetic galaxies are convolved with SDSS and 2MASS bandpasses for
a grid of redshifts in the 0--0.3 range (thick lines). The positions along these
lines that correspond to redshifts of 0, 0.1, 0.2 and 0.3 are marked by squares,
circles, triangles and crosses, respectively, and are connected by thin lines.}
\label{Kpred4}
\end{figure}

Motivated by the surprisingly small residual scatter in the K band flux
prediction (0.1 mag), we have investigated the Bruzual \& Charlot (1993,
2003) model predictions for the distribution of optical and near-IR galaxy
colors. The top left panel in Fig.~\ref{Kpred4} shows that for a given galaxy
type, the $r-K$ color is a function of redshift, and at a given redshift
the $r-K$ color is a function of galaxy type. This behavior is in agreement
with observations, although model galaxies have the $u-r$ colors too blue
by several tenths of a magnitude. The models are also in agreement
with a $J-K$ vs. redshift relation that is independent of galaxy type,
as discussed in Section~\ref{Kpredict} (see top right panel).

The behavior of simple stellar populations (as opposed to synthetic galaxies)
in the same diagrams suggests that both the $r-K$ and $J-K$ colors are
sensitive to metallicity, with about 0.7 mag and 0.2 mag redward shifts
in the $r-K$ and $J-K$ colors as the metallicity increases by a factor of 10.
The observed small residual scatter in the K band flux prediction thus
implies that the galaxy metallicity distribution is fairly narrow: about 0.2 dex
around the median value. This is in agreement with Tremonti et al. (2004)
who found a very tight mass-metallicity relation (0.1 dex metallicity scatter
at a given mass).

\section{Summary and Discussion}

This study indicates the enormous potential of modern massive sensitive large--scale 
surveys, and emphasizes the added value obtained by combining data from different 
wavelengths. While qualitatively our study is in agreement with previous
work (e.g., galaxies detected by IRAS tend to be blue), the sample size and
the wealth of measured parameters allowed us to obtain some qualitatively 
and quantitatively new results. 

Galaxy SEDs form a nearly one-dimensional sequence in the optical-to-near-IR range.
For example, the SDSS $u$ and $r$-band data, supplemented with redshift and 
dust content estimate, can be used to predict $K$-band magnitudes measured by 2MASS 
with an rms scatter of only 0.1 mag and the intrinsic astrophysical scatter probably 
significantly smaller.
Within the restricted wavelength range probed by SDSS, this scatter is even
smaller. Smol\v{c}i\'{c} et al. (2006) show that the rest-frame $r-i$ color 
(5500--8500 \AA\ wavelength range) can be predicted with an rms of only 0.05 mag using 
Str\"omgren colors evaluated in the 4000--5800 \AA\ wavelength range.  Smol\v{c}i\'{c} et al. (2006) also 
find a strong correlation between Str\"omgren colors and the position of a 
galaxy in the BPT diagram, a result that is confirmed here using the $u-r$ color.
As shown by Smol\v{c}i\'{c} et al., other parameters, such as those determined by
Kauffmann et al. (2003a), can also be used to parametrize the position of a galaxy
on this one-dimensional sequence in the multi-dimensional color space. Or equivalently, 
using the terminology from Yip et al. (2004), most of the variance in galaxy SEDs 
is already absorbed in the first few principal components.

We integrate the broad-band UV-to-near-IR SEDs of dominant galaxy types and 
find that the $z$-band flux is the closest, color-independent, proxy for bolometric
flux measurement in the 0.2--2.2 $\mic$ range for galaxies with redshifts smaller 
than $\sim$0.2.

We find that galaxies detected by GALEX include a non-negligible fraction (10-30\%) 
of AGNs, and hence do not represent a clean sample of starburst galaxies. This conclusion
is supported by their $u-r$ color distribution, position in the BPT diagram,
and morphological appearance (see Ag\"{u}eros et al. 2005 for more details).

We demonstrate that interstellar dust content inferred from optical spectra 
by Kauffmann et al. (2003a) is indeed higher for galaxies detected by
IRAS confirming the reliability of the $A_z$ measurement.
This represents a dramatic {\it independent} support for the notion that these 
model-based {\it optical} estimates of $A_z$ are related to the galaxy dust 
content. Furthermore, $A_z$ can be used with the $u$-band measurements to predict 
IRAS 60 $\mu$m flux within a factor of $\sim$2, which suggests that the Bruzual \& 
Charlot (1993, 2003) models used to derive $A_z$ provide fairly good description 
of the relevant physics. Even more detailed and robust analysis 
along these lines will be possible with the advent of GALEX and Spitzer data. 
  
We find that star-forming galaxies tend to be bluer than AGN galaxies for all 
photometric bands bluewards from the $K$-band, while they have redder far-IR-to-optical 
colors. We conclude that star-forming galaxies have redder far-IR-to-optical colors 
than AGN galaxies because they have more UV light that is processed to
far-IR range, as a fraction of the 0.2--2.2 $\mic$
bolometric flux. We emphasize that a variety of different 
data for the {\it same} galaxy sample was required to reach this conclusion: GALEX,
SDSS, 2MASS, and IRAS photometry, as well as the interstellar dust content inferred 
from SDSS optical spectra with  the aid of sophisticated models.

A large sample of galaxies that have SDSS spectra and are detected
by IRAS and NVSS allowed us to study the IR-radio correlation separately 
for star-forming and AGN galaxies. We confirm that both galaxy types follow a tight 
correlation, and find that a large fraction (80\%) of optically classified 
SDSS-NVSS AGN galaxies show significantly more radio emission than expected 
from their IR flux (technically, we show that IR fluxes predicted from observed
radio emission for galaxies that are not detected by IRAS are higher than the
corresponding IRAS upper flux limits, see Section~\ref{radioIR}). 
We also find marginal evidence for different slopes of IR-radio correlation for 
AGN and star-forming subsamples, an effect that seems to be related to the 
$H_\alpha/H_\beta$ line strength ratio.

Perhaps the most important conclusion of our study is that little more than a single
datum can be learned about galaxies from photometric data whose accuracy is not
demonstrably better than $\sim$0.1 mag. Fortunately, all three major modern galaxy
surveys, GALEX, SDSS, and 2MASS, appear to have achieved this goal and, together
with surveys such as FIRST and NVSS, opened unprecedented opportunities for detailed 
studies of galaxies.

\section*{Acknowledgments}

M.O. is grateful to Alex Szalay, Mercedes Filho, and Peter Barthel for
their insights, and to Princeton University and University of Washington
for financial support. 

Funding for the SDSS and SDSS-II has been provided by the Alfred
P. Sloan Foundation, the Participating Institutions, the National
Science Foundation, the U.S. Department of Energy, the National
Aeronautics and Space Administration, the Japanese Monbukagakusho, the
Max Planck Society, and the Higher Education Funding Council for
England. The SDSS Web Site is http://www.sdss.org/.

The SDSS is managed by the Astrophysical Research Consortium for the
Participating Institutions. The Participating Institutions are the
American Museum of Natural History, Astrophysical Institute Potsdam,
University of Basel, Cambridge University, Case Western Reserve
University, University of Chicago, Drexel University, Fermilab, the
Institute for Advanced Study, the Japan Participation Group, Johns
Hopkins University, the Joint Institute for Nuclear Astrophysics, the
Kavli Institute for Particle Astrophysics and Cosmology, the Korean
Scientist Group, the Chinese Academy of Sciences (LAMOST), Los Alamos
National Laboratory, the Max-Planck-Institute for Astronomy (MPA), the
Max-Planck-Institute for Astrophysics (MPIA), New Mexico State
University, Ohio State University, University of Pittsburgh, University
of Portsmouth, Princeton University, the United States Naval
Observatory, and the University of Washington.

The FIRST Survey is supported in part under the auspices of the Department of 
Energy by Lawrence Livermore National Laboratory under contract No. W-7405-ENG-48 
and the Institute for Geophysics and Planetary Physics.

This publication makes use of data products from the Two Micron All Sky Survey, 
which is a joint project of the University of Massachusetts and the Infrared Processing 
and Analysis Center/California Institute of Technology, funded by the National Aeronautics 
and Space Administration and the National Science Foundation.

The Galaxy Evolution Explorer (GALEX) is a NASA Small Explorer. The
mission was developed in cooperation with the Centre National d'Etudes
Spatiales of France and the Korean Ministry of Science and Technology.

\appendix 
\section{Technical summary of the analyzed surveys}

\subsection{Sloan Digital Sky Survey}

SDSS (www.sdss.org) is a digital photometric and spectroscopic survey that 
will cover one quarter of the Celestial Sphere in the North Galactic cap and 
produce a smaller area ($\sim$ 225 deg$^2$), but much deeper, survey in the 
Southern Galactic hemisphere (York et al. 2000, Stoughton et al. 2002, Abazajian
et al. 2003, and references therein). The flux densities of detected objects are 
measured almost simultaneously in five bands ($u$, $g$, $r$, $i$, and $z$, Fukugita
et al. 1996, Hogg et al. 2002, Smith et al. 2002;) with effective wavelengths of 
3551 \AA, 4686 \AA, 6166 \AA, 7480 \AA, and 8932 \AA\ (Gunn et al. 1998),
accurate to 0.02 mag (root-mean-square scatter for sources not limited by photon 
statistics, and also for zeropoints, Ivezi\'{c} et al.~2004b). The survey will 
result in photometric measurements for close to 100 million 
stars and a similar number of galaxies. Astrometric positions are accurate to better than 
0.1 arcsec per coordinate (rms) for sources with $r<20.5^m$ (Pier et al.~2003), and 
the morphological information from the images allows reliable star-galaxy separation 
to $r \sim$ 21.5 mag (Lupton et al.~2002). The imaging data are used to select
sources for follow-up spectroscopic observations, which will result in over a million
spectra. The spectra have a resolution of 1800-2000 in the wavelength range from 3800 
to 9200 \AA. Extragalactic sources targeted in the SDSS spectroscopic survey include 
a flux-limited ``main" galaxy sample ($r$$<$17.77, Strauss {\em et al.} 2002), the 
luminous red galaxy sample (Eisenstein {\em et al.} 2002), and quasars (Richards 
{\em et al.} 2002).

\subsection{ROSAT Survey}
The ROentgen SATellit (ROSAT, 1990-1999) was an X-ray observatory which included
the X-Ray Telescope (XRT) with its 2.4 m focal-length mirror assembly consisting of
four nested Wolter-I mirrors. The focal plane instrumentation consisted of the Position
Sensitive Proportional Counter (PSPC) and the High Resolution Imager (HRI). The Wide-Field
Camera (WFC) with its 0.525 m focal-length mirror assembly consisting of three nested 
Wolter-Schwarzschild mirrors (co-aligned with the XRT). XRT covered $\sim$6-100\AA  
($\sim$2.4-0.12keV) band, and the WFC covered the $\sim$60-300\AA\ ($\sim$0.21-0.05keV) band. 
ROSAT provided a $\sim$2 degree diameter field of view with the PSPC in the focal plane,
and $\sim$40 arcmin diameter field of view with the HRI in the focal plane. The main aim
of the ROSAT mission was the first all-sky survey with imaging X-ray and XUV telescopes;
its X-ray sensitivity was about a factor of 1000 higher than that of the UHURU
satellite. About 100,000 sources have been detected in the survey, an order of magnitude 
more than were known before ROSAT (Voges et al. 1999, Voges et al. 2000). 

\subsection{GALEX Survey}
The Galaxy Evolution Explorer (GALEX) was launched in April 2003, and will 
eventually map the entire sky 
in two bands: the near ultraviolet (NUV; 1750--2800 ${\rm \AA}$) and the far 
ultraviolet (FUV; 1350--1750 ${\rm \AA}$), and to faint flux levels (m=20-25, AB). 
GALEX's 0.5 m telescope and $1.2^{\rm o}$ field of view will also be used to 
make deep observations ($>$tens of kiloseconds) of individual interesting fields 
(such as the Lockman Hole and the Chandra Deep Field--South). The mission's primary 
science goal is to observe star-forming galaxies and to track galaxy evolution.
The GALEX Early Release Observations used here include 
three AIS fields (see www.galex.caltech.edu) which overlap with the SDSS footprint.

\subsection{2MASS Survey}
2MASS used two 1.3-meter telescopes, one at Mt. Hopkins, AZ, and one at CTIO, 
Chile, to survey the entire sky in near-infrared light (see www.ipac.caltech.edu/2mass).  
 Each telescope's camera was equipped with three $256\times256$ arrays (the pixel 
size is 2 arcsec) of HgCdTe detectors which simultaneously observed in the $J$ (1.25 
$\mu{\rm m}$), $H$ (1.65 $\mu{\rm m}$), and $K_s$ (2.17 $\mu{\rm m}$) bands. The detectors 
were sensitive to point sources brighter than about 1 mJy at the $10\sigma$ level, 
corresponding to limiting (Vega-based) magnitudes of 15.8, 15.1, and 14.3, respectively.  
Point-source photometry is repeatable to better than 10\% precision at this level, and 
the astrometric uncertainty for these sources is less than 0.2 arcsec. The 2MASS
catalogs contain positional and photometric information for 470,992,970 point sources 
(2MASS PSC) and 1,647,599 extended sources (2MASS XSC). Details about 2MASS photometry
of galaxies can be found in Jarrett et al. (2000).

\subsection[IRAS]{IRAS Survey}
The Infrared Astronomical Satellite (IRAS, Beichman et al. 1985) produced an 
almost all-sky survey ($\sim$ 98\% of the sky) at 12, 25, 60 and 100 
$\mu{\rm m}$, with the resulting IRAS point source catalog (IRAS PSC) 
containing over 250,000 sources, and the Faint Source Catalog additional
173,000 sources. While the IRAS faint limits are of order 1 Jy, 
it remains a valuable resource due to its important 
wavelength range and nearly full sky coverege.

\subsection{ Radio Surveys }

The basic properties of the radio surveys considered here are:

{\bf GB6:} The Green Bank GB6 survey (GB6, Gregory et al. 1996) is at 4850 MHz (6 cm), 
with 3 arcmin resolution, and covers the declination band between $0^\circ$
and $75^\circ$. The completeness limit of the GB6 catalog is 18 mJy,  
and it includes 75,000 sources.

{\bf NVSS:} The NRAO VLA Sky Survey (NVSS, Condon et al. 1998) is a 1.4 GHz (20 cm) survey with
45 arcsec resolution, and covers the sky north of $-40^\circ$ declination.
The completeness limit of the NVSS catalog is about 2.5 mJy, and it 
includes  1.8 million sources.

{\bf FIRST:} The Faint Images of the Radio Sky at Twenty-centimeters
(FIRST, Becker, White \&  Helfand 1995) is a 1.4 GHz (20 cm) survey with
5 arcsec resolution, and will cover a quarter of the sky
matched to the SDSS footprint. The completeness limit of the FIRST 
catalog is 1 mJy, and it will include about 1 million sources. The FIRST survey 
provides the highest resolution and most accurate radio positions among the 
large radio surveys. It also has the highest source density of about 90 deg$^{-2}$.

{\bf WENSS:} The Westerbork Northern Sky Survey (WENSS, Rengelink et al. 1997) is 
a 326 MHz (92 cm) survey with $54'' \times 54''$cos($\delta$) resolution, and covers the sky
north of $+30^\circ$ declination. The completeness limit of the WENSS catalog 
is 18 mJy, and it includes 230,000 sources.  


\begin{thebibliography}{99}

\bibitem[]{} Abazajian, K., Adelman, J.K., Ag\"{u}eros, M.,  et al. 2003, AJ, 126, 2081
\bibitem[]{} Ag\"{u}eros, M.A., Ivezi\'{c}, \v{Z}., Covey, K.R., et al. 2005, AJ, 130, 1022
\bibitem[]{} Anderson, S.F., Voges, W., Margon, B., et al. 2003, AJ, 126, 2209 
\bibitem[]{} Baldry, I.K, Glazebrook, K., Brinkmann, J., et al. 2004, ApJ, 600, 681
\bibitem[]{} Baldwin, J., Phillips, M., \& Terlevich, R. 1981, PASP, 93, 5
\bibitem[]{} Becker, R.H., White, R.L., \& Helfand, D.J. 1995, ApJ, 450, 559
\bibitem[]{} Beichman, C.A., Neugebauer, G., Habing, H.J., Clegg, P.E.
               \& Chester, T.J. 1985, {\em IRAS Catalogs and Atlases} (US GPO, Washington, DC)
\bibitem[]{} Bell, E.F. 2003, ApJ, 586, 794
\bibitem[]{} Bell, E.F., McIntosh, D.H., Katz, N. \& Weinberg, M.D. 2003, ApJ, 585, L117
\bibitem[]{} Best, P.N., Kauffmann, G., Heckman, T.M. \& Ivezi\'{c}, \v{Z}. 2005a, MNRAS, 362, 9
\bibitem[]{} Best, P.N., Kauffmann, G., Heckman, T.M., et al. 2005b, MNRAS, 362, 25
\bibitem[]{} Blanton, M.R., Hogg, D.W., Bahcall, N.A., et al. 2003, ApJ, 592, 819
\bibitem[]{} Brinchmann, J., Charlot, S., White, S.D.M., Tremonti, C., Kauffmann, G., Heckman, T., Brinkmann, J. 2004, MNRAS, 351, 1151
\bibitem[]{} Bruzual, G. \& Charlot, S. 1993, ApJ, 405, 538
\bibitem[]{} Bruzual, G. \& Charlot, S. 2003, MNRAS, 344, 1000
\bibitem[]{} Buat, V., Iglesias-P\'{a}ramo, J., Seibert, M., et al. 2005, ApJ, 619, 51
\bibitem[]{} Cardelli, J.A., Clayton, G.C., \& Mathis, J.S. 1989, ApJ, 345, 245
\bibitem[]{} Chang, R., Gallazzi, A., Kauffmann, G., Charlot, S., Ivezi\'{c}, \v{Z}., Brinchmann, J. \& Heckman, T.M. 2005,
        accepted by MNRAS (also astro-ph/0502117)
\bibitem[]{} Condon, J.J., Cotton, W.D., Greisen, E.W., Yin, Q.F., 
              Perley, R.A., Taylor, G.B., \& Broderick, J.J. 1998, AJ, 115, 1693
\bibitem[]{} Condon, J.J. \& Broderick, J.J. 1988, AJ, 96, 30
\bibitem[]{} Eisenstein, D.J., Hogg, D.W., Fukugita, M., et al. 2002, ApJ, 585, 694
\bibitem[]{} Finlator, K., Ivezi\'{c}, \v{Z}., Fan, X., et al. 2000, AJ, 120, 2615
\bibitem[]{} Fukugita, M., Ichikawa, T., Gunn, J.E., Doi, M., Shimasaku, K., \& Schneider, 
             D.P. 1996, AJ, 111, 1748
\bibitem[]{} Goto, T. 2005, MNRAS, 360, 322
\bibitem[]{} Gregory, P.C., Scott, W.K., Douglas, K., \& Condon, J.J. 1996, ApJS, 103, 427
\bibitem[]{} Gunn, J.E. \& Oke, J.B. 1975, ApJ, 195, 255
\bibitem[]{} Gunn, J.E., Carr, M., Rockosi, C., et al. 1998, AJ, 116, 3040
\bibitem[]{} Hao, L., Strauss, M.A., Tremonti, C.A., et al. 2005, AJ, 129, 1783
\bibitem[]{} Heckman, T.M., Kauffmann, G., Brinchmann, J., Charlot, S., Tremonti, C. \& White, S.D.M. 2004, ApJ, 613, 109
\bibitem[]{} Helou, G., Soifer, B.T.\& Rowan-Robinson, M. 1985, ApJ, 298, 7 
\bibitem[]{} Hogg, D.W., Finkbeiner, D.P., Schlegel, D.J. \& Gunn, J.E. 2002, AJ, 122, 2129
\bibitem[]{} Hopkins, A.M., Miller, C.J., Nichol, R.C., et al. 2003, ApJ, 599, 971
\bibitem[]{} Ivezi\'{c}, \v{Z}. \& Elitzur, M. 1997, MNRAS 287, 799
\bibitem[]{} Ivezi\'{c}, \v{Z}., Becker, R.H., Blanton, M. et al. 2001a, in AGN Surveys, 
               IAU Colloq.\ 184, eds. R.F.\ Green, E.Ye.\ Khachikian \& D.B. Sanders (San  
               Francisco: ASP), p.\ 137 (astro-ph/0111024)
\bibitem[]{} Ivezi\'{c}, \v{Z}., Tabachnik, S., Rafikov, R., et al. 2001b, AJ, 122, 2749
\bibitem[]{} Ivezi\'{c}, \v{Z}., Menou, K., Knapp, G.R., et al. 2002, AJ, 124, 2364
\bibitem[]{} Ivezi\'{c}, \v{Z}., Siverd, R., Steinhardt, W., et
  al. 2004a, in Multiwavelength AGN Surveys, Guillermo Haro Conf.,
  eds. M\'{u}jica, R. \&  Maiolino, R. (Conzumel, Mexico), p. 53 (astro-ph/0403314)
\bibitem[]{} Ivezi\'{c}, \v{Z}., Lupton, R.H., Schlegel, D., et al. 2004b, AN, 325,  No. 6-8, 583 (astro-ph/0410195)
\bibitem[]{} Jarrett, T.H., Chester, T., Cutri, R., Schneider, S., Skrutskie, M., \& Huchra, 
              J.P. 2000, AJ, 119, 2498
\bibitem[]{} Lupton, R.H., Ivezi\'{c}, \v{Z}., Gunn, J.E., Knapp, G.R., 
             Strauss, M.A. \& Yasuda, N. 2002, in ``Survey and Other Telescope Technologies and 
             Discoveries'', eds. Tyson, J.A. \& Wolff, S., Proceedings of the SPIE, 4836, 350 
\bibitem[]{} Kauffmann, G., Heckman, T.M., White, S.D.M., et al. 2003a, MNRAS, 341, 33
\bibitem[]{} Kauffmann, G., Heckman, T.M., White, S.D.M., et al. 2003b, MNRAS, 341, 54
\bibitem[]{} Kauffmann, G., Heckman, T.M., Tremonti, C.A., et al. 2003c, MNRAS, 346, 1055
\bibitem[]{} Kewley, L., Jansen, R.A. \& Geller, M.J. 2005, PASP 117, 227
\bibitem[]{} McIntosh, D.H., Bell, E.F., Weinberg, M.D. \& Katz, N. 2005, astro-ph/0511737 
\bibitem[]{} Moustakas, J., Kennicutt, R.C. \& Tremonti, C.A. 2005, astro-ph/0511730
\bibitem[]{} Nenkova, M., Ivezi\'{c}, \v{Z}., Elitzur, M. 2002, ApJ, 570, L9
\bibitem[]{} Pasquali, A., Kauffmann, G. \& Heckman, T.M. 2005, MNRAS, 361, 1121
\bibitem[]{} Petrosian, V. 1976, AJ, 209, L1
\bibitem[]{} Pier, J.R., Munn, J.A., Hindsley, R.B., Hennesy, G.S., Kent, S.M., 
             Lupton, R.H. \& Ivezi\'{c}, \v{Z}. 2003, AJ, 125, 1559
\bibitem[]{} Rengelink, R.B., Tang, Y., de Bruyn, A.G., Miley, G.K., Bremer, M.N., Roettgering, H.J.A. 
               \& Bremer, M.A.R. 1997, A\&AS, 124, 259
\bibitem[]{} Richards, G.T., Fan, X., Newberg, H.J., et al. 2002, AJ, 123, 2945
\bibitem[]{} Schlegel, D.J., Finkbeiner, D.P.; Davis, M. 1998, ApJ, 500, 525
\bibitem[]{} Schmitt, H.R., Kinney, A.L., Calzetti, D. \& Storchi Bergman, T. 1997, AJ, 114, 592
\bibitem[]{} Shimasaku, K., Fukugita, M., Doi, M., et al. 2001, AJ, 122, 1238
\bibitem[]{} Smith, J.A., Tucker, D.L., Kent, S.M., et al. 2002, AJ, 123, 2121
\bibitem[]{} Smol\v{c}i\'{c}, V., Ivezi\'{c}, \v{Z}., Ga\'{c}e\v{s}a, M., et al. 2006, submitted to MNRAS
\bibitem[]{} Stoughton, C., Lupton, R.H., Bernardi, M., et al. 2002, AJ, 123, 485
\bibitem[]{} Strateva, I., Ivezi\'{c}, \v{Z}., Knapp, G.R., et al. 2001, AJ, 122, 1861
\bibitem[]{} Strauss, M.A., Weinberg. D.H., Lupton, R.H., et al. 2002, AJ, 124, 1810
\bibitem[]{} Tremonti, C.A., Heckman, T.M., Kauffmann, G., Brinchmann,
  J., Charlot, S., White, S.D.M., Seibert, M., Peng, E.W. Schlegel,
  D.J., Uomoto, A., Fukugita, M. \& Brinkmann, J. 2004, ApJ, 613, 898
\bibitem[]{} Voges, W.H., Aschenbach, B., Boller, Th., et al. 1999, A\&A, 349, 389
\bibitem[]{} Voges, W.H., Aschenbach, B., Boller, Th., et al. 2000, IAU Circ., 7432, 1
\bibitem[]{} van der Kruit, P. 1971, A\&A, 15, 110
\bibitem[]{} Yasuda, N., Fukugita, M., Narayanan, V.K., et al. 2001, AJ, 122, 1104
\bibitem[]{} Yip, C.W., Connolly, A.J., Szalay, A. et al. 2004, AJ, 128, 585
\bibitem[]{} York, D.G., Adelman, J., Anderson, S., et al. 2000, AJ, 120, 1579 
\bibitem[]{} Yun, M.S., Reddy, N.A. \& Condon, J.J. 2001, ApJ, 554, 803
\end{thebibliography}
\end{document}